\documentclass[aps,pre,groupedaddress,twocolumn]{revtex4-2}

\usepackage{amsmath}
\usepackage{amssymb}
\usepackage{graphicx}
\usepackage{color}
\usepackage{bm}
\usepackage{multirow}
\usepackage{latexsym}
\usepackage{comment}

\begin{document}
	
\title{Dependence of protein-induced lipid bilayer deformations on protein shape}

\author{Carlos D. Alas}

\author{Christoph A. Haselwandter}

\affiliation{Department of Physics and Astronomy and Department of Quantitative and Computational Biology, University of Southern California, Los Angeles, CA 90089, USA}

\received{xx} 
\revised{xx} 
\accepted{xx} 

\begin{abstract}

Membrane proteins typically deform the surrounding lipid bilayer membrane, which can play an important role in the function, regulation, and organization of membrane proteins. Membrane elasticity theory provides a beautiful description of protein-induced lipid bilayer deformations, in which all physical parameters can be directly determined from experiments. Analytic treatments of the membrane elasticity theory of protein-induced lipid bilayer deformations have largely focused on idealized protein shapes with circular cross section, and on perturbative solutions for proteins with non-circular cross section. We develop here a boundary value method (BVM) that permits the construction of non-perturbative analytic solutions of protein-induced lipid bilayer deformations for non-circular protein cross sections, for constant as well as variable boundary conditions along the bilayer-protein interface. We apply this BVM to protein-induced lipid bilayer thickness deformations. Our BVM reproduces available analytic solutions for proteins with circular cross section and yields, for proteins with non-circular cross section, excellent agreement with numerical, finite element solutions. On this basis, we formulate a simple analytic approximation of the bilayer thickness deformation energy associated with general protein shapes and show that, for modest deviations from rotational symmetry, this analytic approximation is in good agreement with BVM solutions. Using the BVM, we survey the dependence of protein-induced lipid bilayer thickness deformations on protein shape, and thus explore how the coupling of protein shape and bilayer thickness deformations affects protein oligomerization and transitions in protein conformational state.

\end{abstract}


\maketitle

\section{Introduction}
\label{secIntro}

Membrane proteins play an important role in many essential biological processes, such as signaling, cell shape regulation, and the exchange of molecules between the interior and exterior of cells or between intracellular compartments. Membrane proteins spanning the lipid bilayer are characterized by large hydrophobic regions that approximately match up with the thickness of the lipid bilayer hydrophobic core \cite{engelman05,vanmeer08,Phillips_2013,yeagle_2016,sezgin17,Sych_Mely_Romer_2018}. However, distinct membrane proteins often show distinct hydrophobic thicknesses, and transitions in protein conformational state can change the protein's hydrophobic thickness. Moreover, the lipid composition in cell membranes tends to be highly heterogeneous, with distinct lipids often showing distinct unperturbed lipid bilayer thicknesses. As a result, membrane proteins are generally expected to show a (modest) hydrophobic mismatch with the surrounding lipid bilayer, resulting in protein-induced lipid bilayer thickness deformations \cite{Andersen_Koeppe_2007,Phillips_Ursell_Wiggins_Sens_2009,Huang_1986,Dan_Pincus_Safran_1993,Dan_Berman_Pincus_Safran_1994,Wiggins_Phillips_2005,Ursell_Kondev_Reeves_Wiggins_Phillips_2008,mondal11,Kahraman_Koch_Klug_Haselwandter_PRE_2016,Argudo_Bethel_Marcoline_Wolgemuth_Grabe_2017}. Membrane proteins may also curve the lipid bilayer membrane without perturbing the lipid bilayer thickness \cite{canham70,Helfrich_1973,evans74,zimmerberg06,weikl18,young22}. The energy cost of such protein-induced lipid bilayer deformations depends on the protein shape and conformational state, the lipid composition, membrane mechanical properties such as membrane tension, as well as membrane organization, and can thus regulate, or even determine, membrane protein function.

Membrane elasticity theory provides a beautiful framework for the quantitative description of protein-induced lipid bilayer deformations with, at least in the most basic models, all physical parameters being determined directly from experiments \cite{Andersen_Koeppe_2007,Phillips_Ursell_Wiggins_Sens_2009,Huang_1986,Dan_Pincus_Safran_1993,Dan_Berman_Pincus_Safran_1994,Wiggins_Phillips_2005,Ursell_Kondev_Reeves_Wiggins_Phillips_2008,mondal11,Kahraman_Koch_Klug_Haselwandter_PRE_2016,Argudo_Bethel_Marcoline_Wolgemuth_Grabe_2017,canham70,Helfrich_1973,evans74,zimmerberg06,weikl18,young22,fournier99,Rawicz_Olbrich_McIntosh_Needham_Evans_2000}. As a result, membrane elasticity theory yields definite predictions for the energy cost of protein-induced lipid bilayer deformations and, hence, the coupling between lipid bilayer mechanics and membrane protein function, allowing direct comparisons between theoretical predictions and experimental measurements. Over the past two decades, breakthroughs in membrane protein crystallography and, more recently, cryo-electron microscopy have yielded enormous insight into the shape of membrane proteins. However, analytic treatments of the membrane elasticity theory of protein-induced lipid bilayer deformations have largely focused on idealized protein shapes with a circular cross section, and on perturbative solutions for proteins with non-circular cross section. The objective of this article is to develop, describe, test, and apply a boundary value method (BVM) that permits the construction of non-perturbative analytic solutions of protein-induced lipid bilayer deformations for non-circular protein cross sections. This BVM allows for constant as well as variable boundary conditions along the bilayer-protein interface.

\begin{figure}[t!]
	\center
	\includegraphics[width=\columnwidth]{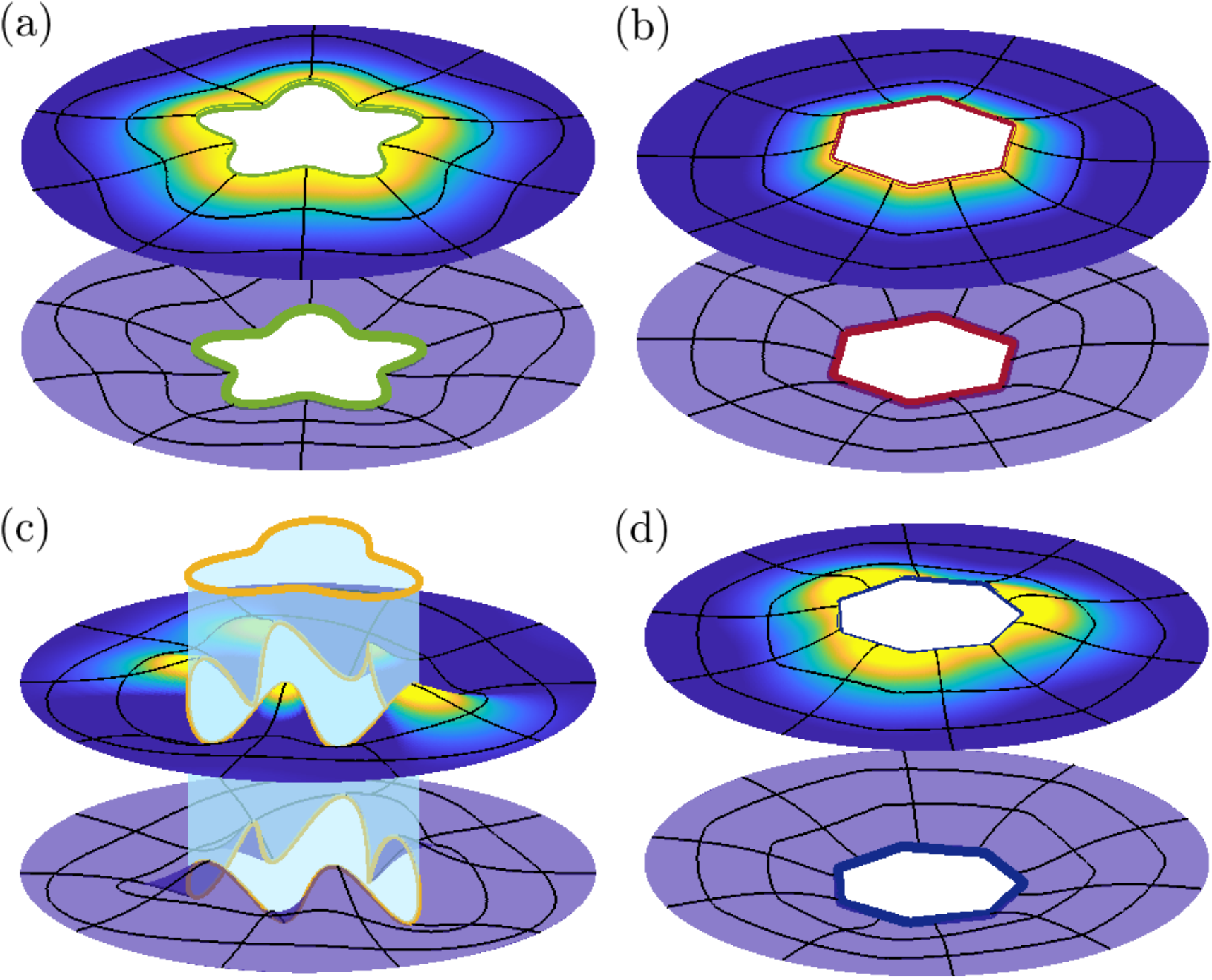}
	\caption{Protein-induced lipid bilayer thickness deformations for selected families of protein shapes: (a) Clover-leaf protein cross section with five-fold symmetry, constant protein hydrophobic thickness, and zero bilayer-protein contact slope, (b) polygonal protein cross section with six-fold symmetry, constant protein hydrophobic thickness, and constant bilayer-protein contact slope $U'=0.3$, (c) clover-leaf protein cross section with three-fold symmetry, a five-fold symmetric (sinusoidal) variation in protein hydrophobic thickness, and zero bilayer-protein contact slope, and (d) polygonal protein cross section with seven-fold symmetry, constant protein hydrophobic thickness, and a three-fold symmetric (sinusoidal) variation in the bilayer-protein contact slope. The color map and purple surfaces show the positions of the upper and lower lipid bilayer leaflets, respectively. The bilayer-protein boundaries are color-coded according to their symmetries (see also Fig.~\ref{fig:3} in Sec.~\ref{secDeformationModel}). For panels (a) and (c) we used $\epsilon=0.2$ and $\epsilon=0.3$ in Eq.~(\ref{eqCdefclover}), respectively, and for panels (b) and (d) we used $P=5$ in Eqs.~(\ref{eqDefCpoly}) and~(\ref{eqDefCpoly2}). All bilayer surfaces were calculated using the reference parameter values in Sec.~\ref{secDeformationModel} and the BVM for protein-induced lipid bilayer thickness deformations described in Sec.~\ref{secBVM}.}
	\label{fig:1}
\end{figure}

In particular, we consider here four generic modes for breaking the rotational symmetry of protein-induced lipid bilayer thickness deformations, which are illustrated in Fig.~\ref{fig:1}. Inspired by observed molecular structures of membrane proteins \cite{vinothkumar2010structures,forrest2015structural}, we consider two classes of non-circular membrane protein cross sections: Clover-leaf [see Fig.~\ref{fig:1}(a)] and polygonal [see Fig.~\ref{fig:1}(b)] protein shapes. Furthermore, we allow for variations in the bilayer-protein hydrophobic mismatch [see Fig.~\ref{fig:1}(c)] as well as in the bilayer-protein contact slope [see Fig.~\ref{fig:1}(d)] along the bilayer-protein interface. Such variations in the bilayer-protein boundary conditions can arise, on the one hand, as inherent features of the protein structure or, on the other hand, as a result of, for instance, the binding of small peptides, such as spider toxins, or other molecules along the bilayer-protein interface \cite{vinothkumar2010structures,forrest2015structural,Phillips_Ursell_Wiggins_Sens_2009,suchyna04}. For each of these four classes of protein shapes we use the BVM to obtain the energy cost of protein-induced lipid bilayer thickness deformations, and test these results against corresponding numerical solutions obtained through the finite element method (FEM) for bilayer thickness deformations \cite{Kahraman_Klug_Haselwandter_2014,Kahraman_Koch_Klug_Haselwandter_SR_2016,Kahraman_Koch_Klug_Haselwandter_PRE_2016}. Inspired by the BVM, we develop a simple analytic approximation of the bilayer thickness deformation energy associated with the general protein shapes illustrated in Fig.~\ref{fig:1}, and investigate the limits of applicability of this analytic approximation. On this basis, we explore how protein shape couples to protein-induced lipid bilayer thickness deformations, and thus affects protein oligomerization and transitions in protein conformational state.

This article is organized as follows. Section~\ref{secDeformationModel} summarizes the elasticity theory of protein-induced lipid bilayer thickness deformations. In Sec.~\ref{secBVM} we describe in detail the BVM for bilayer thickness deformations, test this BVM against FEM solutions, and discuss how the BVM can be used to calculate protein-induced lipid bilayer thickness deformations, and their associated elastic energy, for general protein shapes. On this basis, we develop in Sec.~\ref{secAnalytic} a simple analytic scheme for estimating the energy of protein-induced lipid bilayer thickness deformations for membrane proteins with non-circular cross sections. In Sec.~\ref{secDependShape} we test this analytic approximation against BVM solutions, and survey the dependence of the bilayer thickness deformation energy on membrane protein shape. In Sec.~\ref{secStability} we explore some implications of these results for the self-assembly of protein oligomers and transitions in protein conformational state. A summary and conclusions are provided in Sec.~\ref{secConclusion}.

\section{Modeling protein-induced bilayer thickness deformations}
\label{secDeformationModel}

The preferred hydrophobic thickness of lipid bilayers depends strongly on the lipid chain length \cite{engelman05,vanmeer08,Phillips_2013,yeagle_2016,sezgin17,Sych_Mely_Romer_2018} while different membrane proteins, and even different conformational states of the same membrane protein, often have distinct hydrophobic thicknesses. For membrane proteins that offer a rigid interface to the lipid bilayer, the lipid bilayer thickness tends to deform in the vicinity of the membrane protein so as to achieve hydrophobic matching at the bilayer-protein interface \cite{Andersen_Koeppe_2007,Phillips_Ursell_Wiggins_Sens_2009,Huang_1986,Dan_Pincus_Safran_1993,Dan_Berman_Pincus_Safran_1994,Wiggins_Phillips_2005,Ursell_Kondev_Reeves_Wiggins_Phillips_2008,mondal11,Kahraman_Koch_Klug_Haselwandter_PRE_2016,Argudo_Bethel_Marcoline_Wolgemuth_Grabe_2017}. The resulting protein-induced lipid bilayer thickness deformations can result in a pronounced dependence of protein conformational state, and protein function, on lipid chain length \cite{Huang_1986,mobashery97,Nielsen_Goulian_Andersen_1998,Perozo_Kloda_Cortes_Martinac_2002,Wiggins_Phillips_2004,yuan17}. The purpose of this section is to summarize the elasticity theory of protein-induced lipid bilayer thickness deformations \cite{Andersen_Koeppe_2007,Phillips_Ursell_Wiggins_Sens_2009,Huang_1986,Dan_Pincus_Safran_1993,Dan_Berman_Pincus_Safran_1994,Wiggins_Phillips_2005,Ursell_Kondev_Reeves_Wiggins_Phillips_2008,mondal11,Kahraman_Koch_Klug_Haselwandter_PRE_2016,fournier99,Haselwandter_Phillips_EPL_2013,Haselwandter_Phillips_PLOS_2013}. We first outline the standard elasticity theory of lipid bilayer thickness deformations (see Sec.~\ref{subsecFieldandEnergy}). We then describe how protein shape couples to lipid bilayer thickness, and discuss the models of protein shape considered in this article (see Sec.~\ref{subsecProteinShape}).

\subsection{Elasticity theory of lipid bilayer deformations}
\label{subsecFieldandEnergy}

\begin{figure}[t!]{\includegraphics[width=0.96\columnwidth]{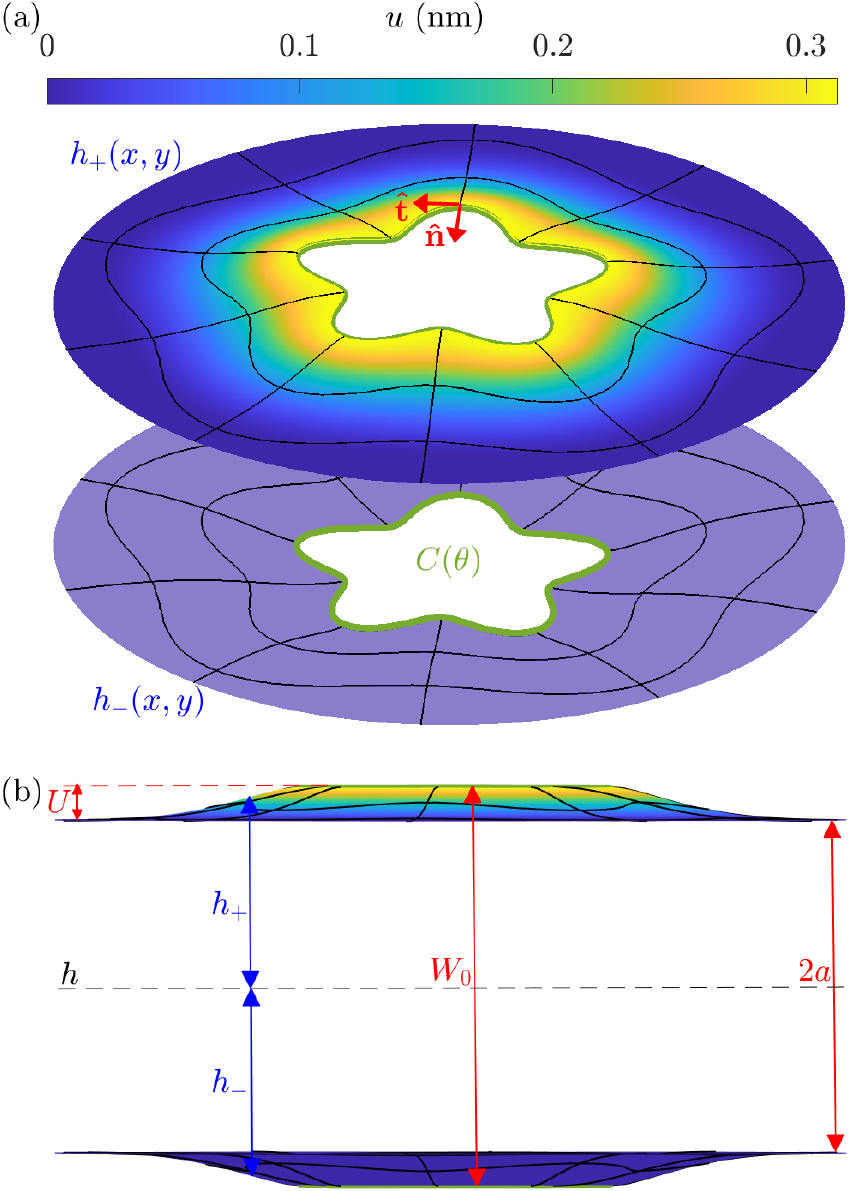}}
	\caption{Notation used for the calculation of protein-induced lipid bilayer thickness deformations in (a) angled and (b) side views. As an example, we consider here a membrane protein with a non-circular (clover-leaf) bilayer-protein boundary curve, $C(\theta)$, constant hydrophobic thickness, $W(\theta)=W_0$, and zero bilayer-protein contact slope, $U^\prime(\theta)=0$. The positions of the upper and lower lipid bilayer leaflets are denoted by $h_+$ and $h_-$, from which the bilayer midplane and bilayer thickness deformation fields $h$ and $u$ can be obtained via Eqs.~(\ref{eqhdef}) and~(\ref{equdef}), respectively. We denote one-half the unperturbed bilayer thickness by $a$, resulting in a hydrophobic mismatch $U=W/2-a$ at the bilayer-protein interface. The unit vectors $\mathbf{\hat{t}}$ and $\mathbf{\hat{n}}$ denote the directions tangential and perpendicular (pointing towards the protein) to the bilayer-protein boundary, respectively.} 
	\label{fig:2}
\end{figure}

Lipid bilayer thickness deformations tend to decay rapidly, with a characteristic decay length $\approx 1$~nm \cite{Ursell_Kondev_Reeves_Wiggins_Phillips_2008,Phillips_Ursell_Wiggins_Sens_2009}. When modeling protein-induced lipid bilayer thickness deformations it is therefore convenient to represent the positions of the two lipid bilayer leaflets in the Monge parameterization of surfaces, $h_\pm=h_\pm(x,y)$, with Cartesian coordinates $(x,y)$ (see Fig.~\ref{fig:2}). It is instructive to express $h_+(x,y)$ and $h_-(x,y)$ in terms of the midplane deformation field $h=h(x,y)$,
\begin{align} \label{eqhdef}
&h = \frac{h_{+} + h_{-}}{2}\,,
\end{align}
and in terms of the thickness deformation field $u=u(x,y)$,
\begin{align} \label{equdef}
&u = \frac{h_{+} - h_{-} - 2a}{2}\,,
\end{align} 
where $a$ is one-half the unperturbed lipid bilayer thickness (Fig.~\ref{fig:2}). The value of $a$ depends on, for instance, the chain length of the lipid species under consideration, and can be directly measured in experiments \cite{Rawicz_Olbrich_McIntosh_Needham_Evans_2000,Andersen_Koeppe_2007,Phillips_Ursell_Wiggins_Sens_2009}.

The membrane elasticity theory describing the shape of lipid bilayers \cite{Seifert_1997,safran03,boal12,Phillips_2013} dates back to the classic work of W. Helfrich \cite{Helfrich_1973}, P.~B. Canham \cite{canham70}, E.~A. Evans \cite{evans74}, and H.~W. Huang \cite{Huang_1986}. Interestingly, one finds that the elastic energies governing $h$ and $u$ in Eqs.~(\ref{eqhdef}) and~(\ref{equdef}) decouple from each other to leading order \cite{fournier99,Wiggins_Phillips_2005}. In the most straightforward model of bilayer-protein interactions 
\cite{Huang_1986,goulian93,Dan_Pincus_Safran_1993,Dan_Berman_Pincus_Safran_1994,Weikl_Kozlov_Helfrich_1998,Andersen_Koeppe_2007,Phillips_Ursell_Wiggins_Sens_2009,boal12,safran03,Seifert_1997,Kahraman_Koch_Klug_Haselwandter_PRE_2016,Wiggins_Phillips_2005,Ursell_Kondev_Reeves_Wiggins_Phillips_2008,mondal11}, the energy cost of protein-induced lipid bilayer midplane deformations is then captured by the classic Helfrich-Canham-Evans energy \cite{canham70,Helfrich_1973,evans74}, and the energy cost of protein-induced lipid bilayer thickness deformations is given by \cite{Huang_1986,Dan_Pincus_Safran_1993,Dan_Berman_Pincus_Safran_1994,Wiggins_Phillips_2005,Ursell_Kondev_Reeves_Wiggins_Phillips_2008,Andersen_Koeppe_2007,Kahraman_Koch_Klug_Haselwandter_PRE_2016,mondal11}
\begin{equation} \label{eqGArea}
G = \frac{1}{2} \int dx dy \left[K_b \left(2 H \right)^2 + K_t \left(\frac{u}{a} \right)^{2} \right]\,,
\end{equation}
where the integral runs over the (in-plane) lipid bilayer surface, $K_b$ is the lipid bilayer bending rigidity, the mean curvature $H=\frac{1}{2}\nabla^{2} u$, and $K_t$ is the bilayer thickness deformation modulus. The terms $K_b \left(\nabla^{2} u \right)^2$ and $K_t \left(u/a \right)^{2}$ in Eq.~(\ref{eqGArea}) provide lowest-order descriptions of the energy cost of bilayer bending and the compression/expansion of the bilayer hydrophobic core, respectively. For simplicity, we assume in Eq.~(\ref{eqGArea}) that the bilayer is under negligible lateral membrane tension. A nonzero membrane tension could be readily included in Eq.~(\ref{eqGArea}) \cite{Haselwandter_Phillips_PLOS_2013,Kahraman_Koch_Klug_Haselwandter_PRE_2016}. We also assume in Eq.~(\ref{eqGArea}) that the lipids forming the bilayer have zero intrinsic curvature. A nonzero lipid intrinsic curvature could also be included in the formalism employed here \cite{Dan_Pincus_Safran_1993,Dan_Berman_Pincus_Safran_1994}.

Similarly as the unperturbed lipid bilayer thickness $2a$, the effective parameters $K_b$ and $K_t$ in Eq.~(\ref{eqGArea}) characterizing the elastic properties of the bilayer membrane depend on the lipid composition, and can be directly measured in experiments \cite{Rawicz_Olbrich_McIntosh_Needham_Evans_2000,Andersen_Koeppe_2007,Phillips_Ursell_Wiggins_Sens_2009}. Typical values of $K_{b}, K_{t}$, and $a$ for cell membranes are $K_{b}=20$~$k_{B}T$, $K_{t}=60$~$k_{B}T$/nm$^{2}$, and $a=1.6$~nm \cite{Niggemann_Kummrow_Helfrich_1995, Rawicz_Olbrich_McIntosh_Needham_Evans_2000, Ursell_Kondev_Reeves_Wiggins_Phillips_2008, Pan_Tristram-Nagle_Kucerka_Nagle_2008}. Unless stated otherwise, we use here these values of $K_{b}, K_{t}$, and $a$. When studying the dependence of protein-induced bilayer thickness deformations on lipid chain length we follow Refs.~\cite{Rawicz_Olbrich_McIntosh_Needham_Evans_2000,Wiggins_Phillips_2005,Shrestha_Kahraman_Haselwandter_2020} and assume, for simplicity, a linear dependence of $a$ on lipid chain length:
\begin{equation} \label{eqDefm}
a = \frac{1}{2} \left(0.13 m + 1.7\right)\,\text{nm}\,.
\end{equation}
The integer $m$ in Eq.~(\ref{eqDefm}) denotes the lipid chain length (number of carbon atoms comprising each lipid chain), with the approximate range $13 \leq m \leq 22$ for phospholipids in cell membranes \cite{engelman05,vanmeer08,sezgin17,Rawicz_Olbrich_McIntosh_Needham_Evans_2000}. For simplicity, we take $K_b$ and $K_t$ to be independent of $m$ while noting that, in general, $K_b$ and $K_t$ may have a (weak) dependence on $m$ \cite{Rawicz_Olbrich_McIntosh_Needham_Evans_2000}.

The effective parameters $K_{b}, K_{t}$, and $a$ in Eq.~(\ref{eqGArea}) yield the characteristic length scale
\begin{equation} \label{eqDeflambda}
\lambda = \left(\frac{a^{2}K_{b}}{K_{t}}\right)^{1/4}\,,
\end{equation}
which corresponds to the characteristic decay length of bilayer thickness deformations \cite{Wiggins_Phillips_2005}. As alluded to above, we have $\lambda \approx 1$~nm \cite{Ursell_Kondev_Reeves_Wiggins_Phillips_2008,Phillips_Ursell_Wiggins_Sens_2009}. Similarly, the bilayer bending rigidity $K_b$ defines a characteristic energy scale in Eq.~(\ref{eqGArea}). It is therefore convenient to recast the bilayer thickness deformation energy in Eq.~(\ref{eqGArea}) in terms of the characteristic spatial and energy scales, $\lambda$ and $K_b$. Unless stated otherwise, we use here a dimensionless form of Eq.~(\ref{eqGArea}) such that $G\rightarrow \bar G K_{b}$, $x\rightarrow \bar x \lambda$, $y\rightarrow \bar y \lambda$, $u\rightarrow \bar u \lambda$, $a\rightarrow \bar a \lambda$, and $K_{t}\rightarrow \bar K_{t} K_b/\lambda^{2}$, resulting in
\begin{equation} \label{eqGAreaDL}
\bar G = \frac{1}{2} \int d\bar x d\bar y \left[\left(\bar \nabla^{2} \bar u \right)^2 + \bar u^{2} \right]\,,
\end{equation}
where $\bar \nabla=\lambda \nabla$.

We assume that, for a given protein conformational state, the dominant bilayer thickness deformation field $\bar u(\bar x,\bar y)$ minimizes Eq.~(\ref{eqGAreaDL}) subject to suitable boundary conditions \cite{Huang_1986,Dan_Pincus_Safran_1993,Dan_Berman_Pincus_Safran_1994,Wiggins_Phillips_2005,Ursell_Kondev_Reeves_Wiggins_Phillips_2008,Andersen_Koeppe_2007,Kahraman_Koch_Klug_Haselwandter_PRE_2016}. The Euler-Lagrange equation associated with Eq.~(\ref{eqGAreaDL}) is given by
\begin{equation} \label{eqHelmholtz}
\left(\bar \nabla^{2} - \bar \nu_{+} \right) \left(\bar \nabla^{2} - \bar \nu_{-} \right)  \bar u = 0
\end{equation}
with $\bar \nu_{\pm} = \pm \mathbf{i}$, where $\mathbf{i}$ is the imaginary unit. To construct the general solution of Eq.~(\ref{eqHelmholtz}) for protein-induced bilayer thickness deformations it is useful to transform $\left(\bar x,\bar y\right)$ to the dimensionless polar coordinates $\left(\bar r,\theta\right)$ with the protein center as the origin of the polar coordinate system. Assuming that protein-induced bilayer thickness deformations are localized about the protein we have $\bar u \to 0$ as $\bar r \to \infty$ \cite{Huang_1986,Nielsen_Goulian_Andersen_1998,Nielsen_Andersen_2000}, in which case Eq.~(\ref{eqHelmholtz}) yields \cite{Haselwandter_Phillips_EPL_2013,Haselwandter_Phillips_PLOS_2013,Kahraman_Koch_Klug_Haselwandter_PRE_2016}
\begin{align} \label{equEigenSuper}
\bar u\left(\bar r,\theta\right) = \bar f^{+}\left(\bar r,\theta\right) + \bar f^{-}\left(\bar r,\theta\right)\,,
\end{align}
where the Fourier-Bessel series
\begin{align} \label{eqEigenFuncs}
\begin{split}
\bar f^{\pm}(\bar r,\theta) = & A^{\pm}_{0}K_{0} \left(\sqrt{\bar \nu_\pm} \bar r \right) + \\
\sum^{\infty}_{n = 1}  \bigg[ A^{\pm}_{n}   K_{n}&\left( \sqrt{\bar \nu_\pm} \bar r\right) \cos\left( n \theta \right) + B^{\pm}_{n} K_{n}\left( \sqrt{\bar \nu_\pm} \bar r\right) \sin\left( n \theta \right) \bigg]\,,
\end{split}
\end{align}
in which the $K_{n}$ are the modified Bessel functions of the second kind \cite{Abramowitz_Stegun_1964} and the values of the coefficients $A^{\pm}_{0}$, $A^{\pm}_{n}$, and $B^{\pm}_{n}$ are determined by the bilayer-protein boundary conditions.

The bilayer thickness deformation energy in Eq.~(\ref{eqGAreaDL}) is conveniently evaluated for the stationary bilayer thickness deformation field in Eq.~(\ref{equEigenSuper}) by noting that, via Eq.~(\ref{eqHelmholtz}), Eq.~(\ref{eqGAreaDL}) can be transformed to a line integral along the bilayer-protein boundary $\bar C$ \cite{Wiggins_Phillips_2005,Haselwandter_Phillips_PLOS_2013,Kahraman_Koch_Klug_Haselwandter_PRE_2016} (Fig.~\ref{fig:2}). For simplicity, we thereby take the bilayer-protein boundary to be specified by the polar curve $\mathbf{\bar r}=\bar C(\theta) \mathbf{\hat{r}}$, where $\mathbf{\hat{r}}$ is the radial unit vector pointing away from the protein center. We thus have
\begin{equation} \label{eqGLine}
\bar G = \frac{1}{2} \int_0^{2 \pi} d \theta \, \bar l \, \mathbf{\hat{n}} \cdot \left( \bar \nabla \bar u \bar \nabla^{2} \bar u - \bar u \bar \nabla^{3} \bar u \right) \big |_{\bar r=\bar C(\theta)}\,,
\end{equation}
where the line element $\bar l=\sqrt{\left[\bar C(\theta)\right]^{2}+ \left[\bar C'(\theta)\right]^{2}}$, and the unit vector $\mathbf{\hat{n}}$ is normal to the tangent of $\mathbf{\bar r}=\bar C(\theta) \mathbf{\hat{r}}$ and points towards the protein (Fig.~\ref{fig:2}). Note that the term in brackets in Eq.~(\ref{eqGLine}) may be interpreted as a bilayer-protein line tension along the bilayer-protein boundary \cite{Wiggins_Phillips_2004,Wiggins_Phillips_2005,Kahraman_Koch_Klug_Haselwandter_PRE_2016}. The normal vector $\mathbf{\hat{n}}$ in Eq.~(\ref{eqGLine}) is obtained by differentiating the bilayer-protein boundary curve $\mathbf{\bar r}=\bar C(\theta) \mathbf{\hat{r}}$ with respect to $\theta$ and rotating the resulting tangent vector by $\pi/2$ so as to point towards the protein,
\begin{equation} \label{eqndef}
\mathbf{ \hat{n} } = \frac{-  \bar C(\theta) \pmb{ \hat{r} } + \bar C'(\theta) \pmb{ \hat{\theta} }  } { \bar l } \,,
\end{equation}
where we have noted that the (counterclockwise) angular unit vector $\pmb{\hat{\theta}}=d \mathbf{\hat{r}}/d\theta$ in polar coordinates (Fig.~\ref{fig:2}). Equation~(\ref{eqGLine}) with Eq.~(\ref{eqndef}) allows calculation of $\bar G$ in Eq.~(\ref{eqGAreaDL}) and, hence, $G$ in Eq.~(\ref{eqGArea}) along a one-dimensional curve rather than over a two-dimensional surface, which provides a computationally efficient method for evaluating~$\bar G$.

\subsection{Modeling protein shape}
\label{subsecProteinShape}

The coefficients $A^{\pm}_{0}$, $A^{\pm}_{n}$, and $B^{\pm}_{n}$ in Eq.~(\ref{eqEigenFuncs}) are fixed by the boundary conditions at the bilayer-protein interface. The general mathematical form of these boundary conditions, which encode the key protein properties governing protein-induced lipid bilayer thickness deformations, follows from the calculus of variations \cite{Hilbert1953,vanbrunt04}. Based on previous work on protein-induced bilayer thickness deformations \cite{Huang_1986,Dan_Pincus_Safran_1993,Dan_Berman_Pincus_Safran_1994,Nielsen_Goulian_Andersen_1998,Andersen_Koeppe_2007,Phillips_Ursell_Wiggins_Sens_2009}, we assume that the lipid bilayer thickness deforms in the vicinity of membrane proteins so as to achieve hydrophobic matching at the bilayer-protein interface. We thus have the boundary condition
\begin{equation} \label{eqDefU}
\bar u(\bar r, \theta) \big|_{\bar r=\bar C(\theta)} = \bar U(\theta)\,,
\end{equation}	
where the bilayer-protein hydrophobic mismatch
\begin{equation} \label{eqUWa}
\bar U(\theta) = \frac{1}{2}\left[\bar W(\theta) - 2 \bar a\right] \,,
\end{equation}
in which $W(\theta) = \lambda \bar W(\theta)$ is the protein hydrophobic thickness along the bilayer-protein boundary (Fig.~\ref{fig:2}). For large enough magnitudes of $U$, membrane proteins or lipids may expose parts of their hydrophobic regions to water, which would amount to an offset of $\bar W$ in Eq.~(\ref{eqUWa}). For a given membrane protein, $W(\theta)$ can be estimated from the molecular structure of the membrane protein \cite{Andersen_Koeppe_2007,Phillips_Ursell_Wiggins_Sens_2009,Haselwandter_Phillips_PLOS_2013,Kahraman_Koch_Klug_Haselwandter_SR_2016,Kahraman_Koch_Klug_Haselwandter_PRE_2016} and/or computer simulations \cite{Elmore_Dougherty_2003,Argudo_Bethel_Marcoline_Wolgemuth_Grabe_2017}. We explore here protein-induced bilayer thickness deformations for generic models of $W(\theta)$ inspired by the molecular structure of the mechanosensitive channel of large conductance (MscL) \cite{Chang_Spencer_Lee_Barclay_Rees_1998,Doyle_1998,Sukharev_Durell_Guy_2001,Perozo_Cortes_Sompornpisut_Kloda_Martinac_2002}.

In addition to Eq.~(\ref{eqDefU}), it is also necessary to specify boundary conditions on the (normal) derivative of $u$ at the bilayer-protein interface \cite{Hilbert1953,vanbrunt04}. Based on Refs.~\cite{Huang_1986,Nielsen_Goulian_Andersen_1998,Argudo_Bethel_Marcoline_Wolgemuth_Grabe_2017,Wiggins_Phillips_2004,Wiggins_Phillips_2005,Ursell_Kondev_Reeves_Wiggins_Phillips_2008} we generally focus on the fixed-value boundary condition
\begin{align} \label{eqDefUp}
\mathbf{ \hat{n} } \cdot \bar \nabla \bar u(\bar r, \theta) \big|_{\bar r=\bar C(\theta)} = \bar U'(\theta)\,,
\end{align}
but also explore choices for $\bar U'(\theta)$ minimizing the bilayer thickness deformation energy. We allow for constant as well as varying $\bar U'(\theta)$ in Eq.~(\ref{eqDefUp}).

For a (hypothetical) membrane protein with a perfectly circular cross section $\bar C(\theta) =\bar R$, where $\bar R$ is the (dimensionless) protein radius, and constant $\bar U$ and $\bar U'$, the bilayer-protein boundary conditions in Eqs.~(\ref{eqDefU}) and~(\ref{eqDefUp}) are azimuthally symmetric about the protein center, and the resulting protein-induced bilayer thickness deformations also show azimuthal symmetry about the protein center \cite{Huang_1986,Nielsen_Goulian_Andersen_1998,Wiggins_Phillips_2004,Wiggins_Phillips_2005}. Equations~(\ref{eqDefU}) and~(\ref{eqDefUp}) suggest three, not mutually exclusive, modes for protein structures to break rotational symmetry, and to hence endow protein-induced bilayer thickness deformations with a non-trivial structure \cite{Haselwandter_Phillips_PLOS_2013,Haselwandter_Phillips_EPL_2013}. First, the value of $\bar U$ in Eq.~(\ref{eqDefU}) or, second, the value of $\bar U'$ in Eq.~(\ref{eqDefUp}) may vary along the bilayer-protein interface. To explore generic effects of varying $\bar U$ or $\bar U'$ on protein-induced bilayer thickness deformations we consider the bilayer-protein hydrophobic mismatch
\begin{equation} \label{eqUBC}
\bar U(\theta) = \bar U_{0} + \bar \beta \cos(w \theta)
\end{equation}
and the bilayer-protein contact slope
\begin{equation} \label{eqUpBC}
\bar U'(\theta) = \bar U'_{0} + \bar \gamma \cos(v \theta)\,,
\end{equation}
where $\bar U_{0}$ and $\bar U'_{0}$ denote the average bilayer-protein hydrophobic mismatch and bilayer-protein contact slope, $\bar \beta$ and $\bar \gamma$ denote the amplitudes of the perturbations about these average values, and $w$ and $v$ denote the protein symmetries associated with variations in $\bar U$ and $\bar U'$. Unless stated otherwise, we set here $\bar U_{0} \lambda=-0.1$~nm and $\bar \beta \lambda=0.5$~nm in Eq.~(\ref{eqUBC}) for all calculations involving a modulation in the bilayer-protein hydrophobic mismatch, and $\bar U'_{0}=0$ and $\bar \gamma=0.3$ in Eq.~(\ref{eqUpBC}) for all calculations involving a modulation in the bilayer-protein contact slope. For all scenarios considered here in which we keep $\bar U$ or $\bar U'$ constant along the bilayer-protein interface we set, unless stated otherwise, $\bar U \lambda=0.3$~nm or $\bar U'=0$. The values of $U$ employed here are in line with previous work on MscL and gramicidin channels \cite{Ursell_Kondev_Reeves_Wiggins_Phillips_2008,Chang_Spencer_Lee_Barclay_Rees_1998,Perozo_Cortes_Sompornpisut_Kloda_Martinac_2002,Argudo_Bethel_Marcoline_Wolgemuth_Grabe_2017,Nielsen_Goulian_Andersen_1998}. Note that the ranges of $\bar U$ and $\bar U'$ considered here satisfy the constraints $|U| < a$ and $|U'| < 1$ underlying the use of the leading-order energy in Eq.~(\ref{eqGArea}) to describe protein-induced bilayer thickness deformations \cite{Huang_1986,Dan_Pincus_Safran_1993,Dan_Berman_Pincus_Safran_1994,fournier99,Wiggins_Phillips_2005,Ursell_Kondev_Reeves_Wiggins_Phillips_2008,Andersen_Koeppe_2007,Kahraman_Koch_Klug_Haselwandter_PRE_2016}.

\begin{figure}[t!]{\includegraphics{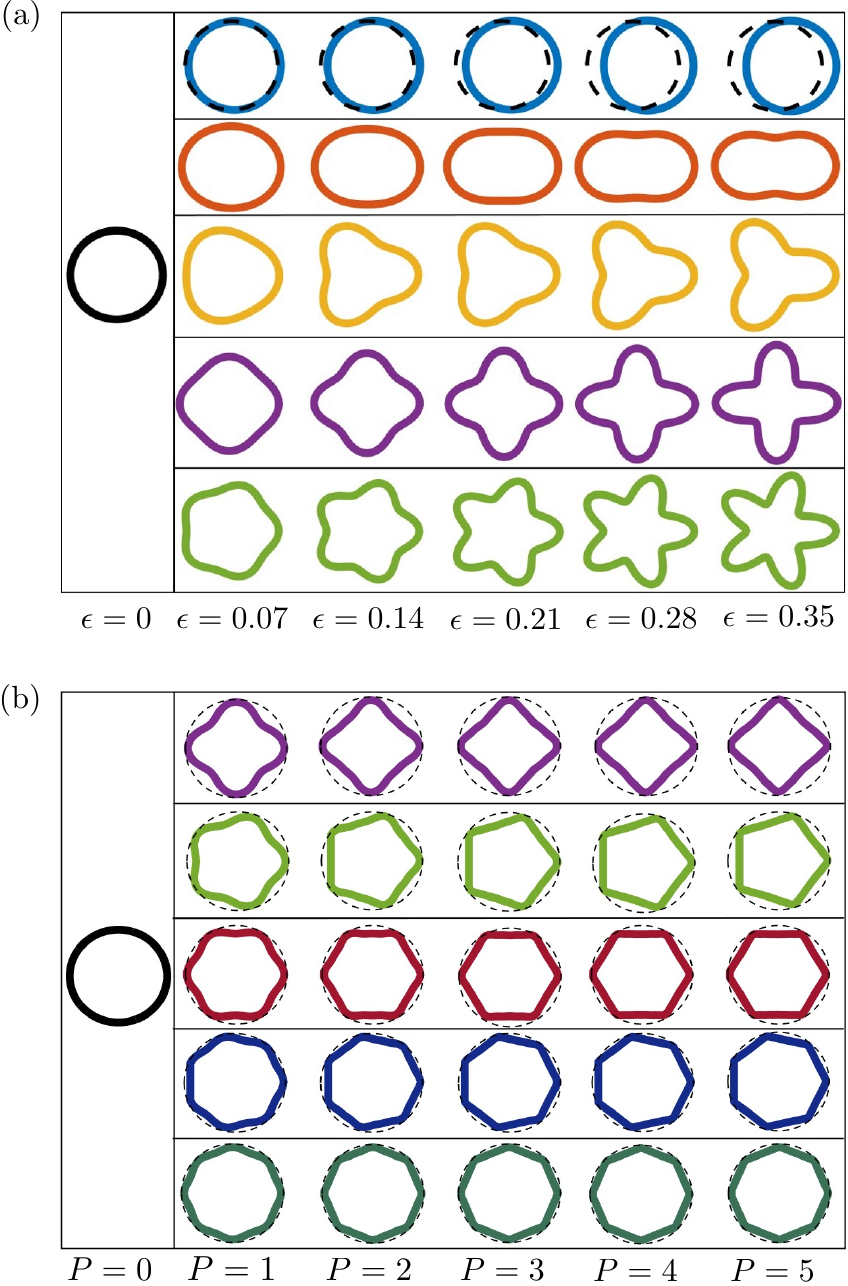}}
	\caption{Cross sections of cylindrical protein shapes (left-most panels) and (a) clover-leaf and (b) polygonal protein shapes (right panels). The clover-leaf protein cross sections in panel (a) are obtained from Eq.~(\ref{eqCdefclover}) with $\epsilon=0.07$, 0.14, 0.21, 0.28, and 0.35 (left to right) and $s=1$, 2, 3, 4, and 5 (top to bottom), with $\epsilon =0$ yielding a circular protein cross section. Note that the clover-leaf protein cross sections with $s=1$ only show small deviations from the corresponding circular protein cross section obtained with $\epsilon = 0$ in Eq.~(\ref{eqCdefclover}) (dashed curves) for the values of $\epsilon$ considered here. The polygonal protein cross sections in panel (b) are obtained from Eq.~(\ref{eqDefCpoly}) with $P=1$, 2, 3, 4, and 5 (left to right) and $s=4$, 5, 6, 7, and 8 (top to bottom). As a guide to the eye, these polygonal protein cross sections are inscribed in circles obtained with $P = 0$ in Eq.~(\ref{eqDefCpoly}) (dashed curves).}
	\label{fig:3}
\end{figure}

Angular variations in $\bar C(\theta)$ along the bilayer-protein boundary $\mathbf{\bar r}=\bar C(\theta) \mathbf{\hat{r}}$ provide, in addition to Eqs.~(\ref{eqUBC}) and~(\ref{eqUpBC}), a third mode for a protein structure to break azimuthal symmetry of protein-induced bilayer thickness deformations about the protein center. Inspired by molecular structures of tetrameric and pentameric MscL \cite{Chang_Spencer_Lee_Barclay_Rees_1998,Liu2009,Haselwandter_Phillips_PLOS_2013,Kahraman_Koch_Klug_Haselwandter_SR_2016} and other membrane proteins \cite{vinothkumar2010structures,forrest2015structural}, we consider here two generic classes of protein shapes breaking rotational symmetry. On the one hand, we consider clover-leaf protein cross sections specified by
\begin{equation} \label{eqCdefclover}
\bar C(\theta) = \bar R \big{[} 1 + \epsilon \cos( s \theta  ) \big{]},
\end{equation}
where $\epsilon$ parameterizes the magnitude of deviations from a circular protein cross section, $\epsilon=0$ for circular protein cross sections, and $s$ denotes the symmetry of the boundary curve [see Figs.~\ref{fig:2}(a) and~\ref{fig:3}(a)]. On the other hand, we consider (rounded) polygonal protein cross sections specified by the series
\begin{align}
\begin{split}
&\bar C(\theta) =\\ 
&\bar A_{\bar R}\sqrt{\Biggr{[}\sum^{P}_{ p = -P }\frac{\cos(sp + 1)\theta}{(sp + 1)^{2}}\Biggr{]}^{2} + \Biggr{[}\sum^{P}_{ p = -P }\frac{\sin(sp + 1)\theta}{(sp + 1)^{2}} \Biggr{]}^{2} }\,,
\end{split} \label{eqDefCpoly}
\end{align}
where larger $P$ yield sharper polygonal corners with $P=0$ for circular protein cross sections, $\bar A_{\bar R}$ is a rescaling factor chosen so as to ensure that the polygons are inscribed in circles of radius $\bar R$, and $s$ denotes the polygonal symmetry \cite{Robert_1994,Haselwandter_Phillips_PLOS_2013} [see Fig.~\ref{fig:3}(b)]. We have
\begin{equation} \label{eqDefCpoly2}
\bar A_{\bar R} = \frac{\bar R}{\sum^{P}_{ p = -P }\frac{1}{(sp + 1)^{2}}}\,.
\end{equation}
Unless stated otherwise, we set $P=5$ in Eqs.~(\ref{eqDefCpoly}) and~(\ref{eqDefCpoly2}). For the scenarios considered here we found that an increase in $P$ beyond $P=5$ only resulted in minor shifts in the bilayer thickness deformation energy. For all the calculations described here we use $\bar R \lambda\approx2.3$~nm in Eqs.~(\ref{eqCdefclover}) and~(\ref{eqDefCpoly2}), which approximately corresponds to the observed size of a closed state of MscL~\cite{Chang_Spencer_Lee_Barclay_Rees_1998,Wiggins_Phillips_2005}.

\section{Boundary value method for bilayer thickness deformations}
\label{secBVM}

In this section we introduce a BVM for bilayer thickness deformations, which allows calculation of protein-induced bilayer thickness deformations, and their associated elastic energy, for general protein shapes. In the following sections we use this BVM to calculate the bilayer thickness deformation field $\bar u(\bar r,\theta)$ in Eq.~(\ref{equEigenSuper}), and the resulting bilayer thickness deformation energy $G$ in Eq.~(\ref{eqGArea}), for the clover-leaf and polygonal protein shapes $\mathbf{\bar r}=\bar C(\theta) \mathbf{\hat{r}}$ in Eqs.~(\ref{eqCdefclover}) and~(\ref{eqDefCpoly}) and the boundary conditions $\bar U(\theta)$ and $\bar U'(\theta)$ in Eqs.~(\ref{eqUBC}) and~(\ref{eqUpBC}). We first provide a general formulation of the BVM for bilayer thickness deformations, and validate this BVM against exact analytic and FEM solutions (see Sec.~\ref{subsecMethod}). We then discuss how the numerical performance of the BVM for bilayer thickness deformations can be improved by employing an adaptive point distribution (APD) that results in a nonuniform distribution of boundary points for non-circular protein cross sections (see Sec.~\ref{subsubsecAPD}).

\subsection{Formulation and validation of the boundary value method}
\label{subsecMethod}

The BVM for bilayer thickness deformations takes the analytic solution for $\bar u(\bar r,\theta)$ in Eq.~(\ref{equEigenSuper}) as its starting point, and assumes that the infinite series in this general solution can be truncated at some finite order $N$:
\begin{align} \label{equEigenSuperN}
\bar u\left(\bar r,\theta\right) \approx \bar f^{+}_N\left(\bar r,\theta\right) + \bar f^{-}_N\left(\bar r,\theta\right)\,,
\end{align}
where
\begin{align} \label{eqEigenFuncsN}
\begin{split}
\bar f^{\pm}_N(\bar r,\theta) = & A^{\pm}_{0}K_{0} \left(\sqrt{\bar \nu_\pm} \bar r \right) + \\
\sum^{N}_{n = 1}  \bigg[ A^{\pm}_{n}   K_{n}&\left( \sqrt{\bar \nu_\pm} \bar r\right) \cos\left( n \theta \right) + B^{\pm}_{n} K_{n}\left( \sqrt{\bar \nu_\pm} \bar r\right) \sin\left( n \theta \right) \bigg]\,.
\end{split}
\end{align}
The solution in Eq.~(\ref{equEigenSuperN}) with Eq.~(\ref{eqEigenFuncsN}) contains the $4N+2$ unknown constants $A^{\pm}_{0}$, $A^{\pm}_{n}$, and $B^{\pm}_{n}$. In the BVM for bilayer thickness deformations, we fix these coefficients by imposing the boundary conditions in Eqs.~(\ref{eqDefU}) and~(\ref{eqDefUp}) at $2N+1$ boundary points along the bilayer-protein interface. For now, we take these boundary points to be uniformly distributed along the bilayer-protein interface, with a constant arc length separating adjacent boundary points along the bilayer-protein interface. We return to the distribution of boundary points in Sec.~\ref{subsubsecAPD}.

From the boundary conditions in Eqs.~(\ref{eqDefU}) and~(\ref{eqDefUp}) we have
\begin{eqnarray} \label{eqBCmU}
\bar u(\bar r, \theta_j) \big|_{\bar r=\bar C\left(\theta_j\right)} &=& \bar U\left(\theta_j\right)\,,\\ \label{eqBCmUp}
\mathbf{ \hat{n} } \cdot \bar \nabla \bar u(\bar r, \theta_j) \big|_{\bar r=\bar C\left(\theta_j\right)} &=& \bar U'\left(\theta_j \right)\,,
\end{eqnarray}
in which $j=1,2,\dots,2N+1$ denote the boundary points along the bilayer-protein interface, where $\bar C(\theta)=\bar R$ for proteins with a circular cross section, $\bar C(\theta)$ is as in Eq.~(\ref{eqCdefclover}) for clover-leaf protein shapes, and $\bar C(\theta)$ is as in Eq.~(\ref{eqDefCpoly}) for polygonal protein shapes (Fig.~\ref{fig:3}). Equations~(\ref{eqBCmU}) and~(\ref{eqBCmUp}) amount to a linear system of equations
\begin{equation} \label{eqmeq}
\mathbf{A} \mathbf{x} = \mathbf{b}\,,
\end{equation}
where the vector $\mathbf{x}$ has dimension $4N+2$ and contains the unknown constants  $A^{\pm}_{0}$, $A^{\pm}_{n}$, and $B^{\pm}_{n}$, the $4N+2$ components of the vector $\mathbf{b}$ contain the boundary conditions on the right-hand sides of Eqs.~(\ref{eqBCmU}) and~(\ref{eqBCmUp}), and $\mathbf{A}$ is a square matrix of order $4N+2$ that has the coefficients of the constants $A^{\pm}_{0}$, $A^{\pm}_{n}$, and $B^{\pm}_{n}$ on the left-hand sides of Eqs.~(\ref{eqBCmU}) and~(\ref{eqBCmUp}) as its entries. Equation~(\ref{eqmeq}) can be solved efficiently using the extensive numerical methods available for the solution of matrix equations. We employed here LU decomposition with partial pivoting to solve Eq.~(\ref{eqmeq}) for $\mathbf{x}$ \cite{Atkinson_1988,Trefethen_1997}.

\begin{figure}[t!]{\includegraphics{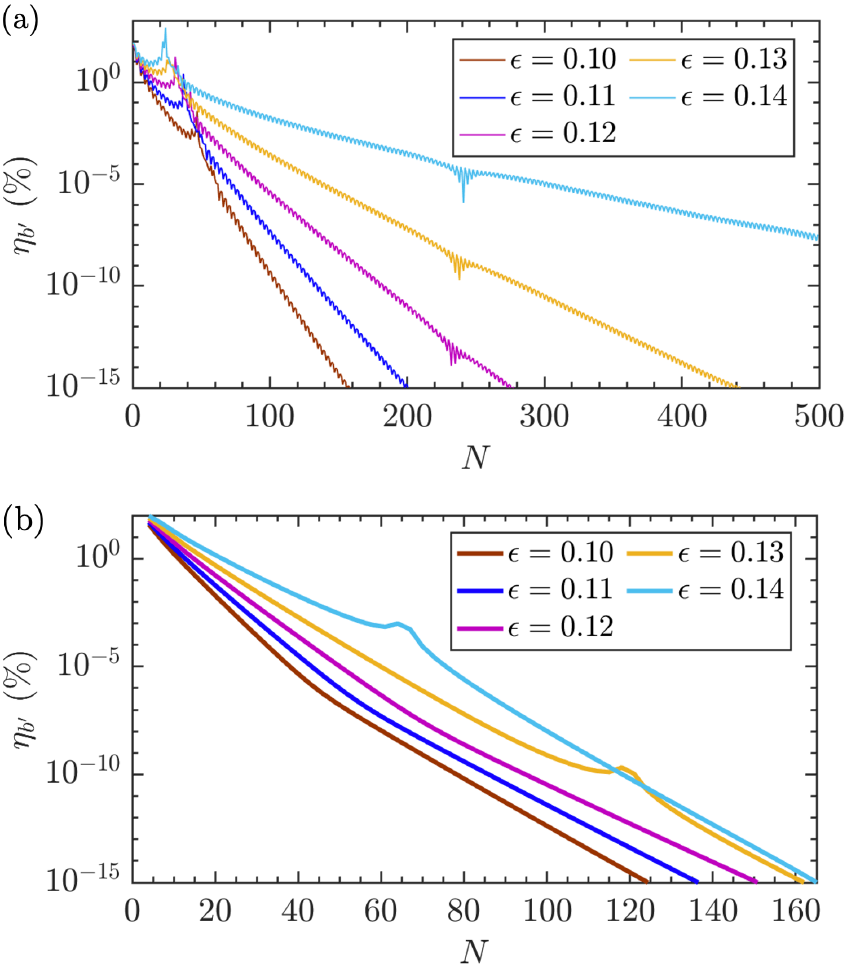}}
	\caption{Percentage difference between the exact bilayer thickness deformation field along the bilayer-protein boundary and the bilayer thickness deformation field obtained from the BVM solution, $\eta_{b'}$ in Eq.~(\ref{eqetabdef}), as a function of the number of terms in the Fourier-Bessel series in Eq.~(\ref{equEigenSuperN}) with Eq.~(\ref{eqEigenFuncsN}) for (a) uniformly distributed points along the bilayer-protein boundary and (b) the boundary point distributions implied by the APD method (see Sec.~\ref{subsubsecAPD}). For both panels we considered three-fold clover-leaf protein shapes ($s=3$) in Eq.~(\ref{eqCdefclover}) with the indicated values of $\epsilon$, $\bar R \lambda\approx2.3$~nm, and the constant $\bar U \lambda=0.3$~nm and $\bar U'=0$. In panel (b) we used, for ease of comparison, the same gap factor $\Omega=0.25$ in Eq.~(\ref{eqelldef}) for all curves.}
	\label{fig:4}
\end{figure}

To quantify numerical errors in our BVM solutions it is useful to compute, based on the calculated $A^{\pm}_{0}$, $A^{\pm}_{n}$, and $B^{\pm}_{n}$ in Eq.~(\ref{equEigenSuperN}) with Eq.~(\ref{eqEigenFuncsN}), the values of $\bar u(\bar r,\theta)$ along the bilayer-protein boundary for a given set of reference points distinct from the boundary points employed for the BVM solution in Eq.~(\ref{eqmeq}). We compile these computed boundary values of $\bar u(\bar r,\theta)$ in a vector $\mathbf{\tilde{b}}$, and compare $\mathbf{\tilde{b}}$ to the corresponding exact boundary values $\mathbf{b'}$ mandated by the boundary conditions in Eqs.~(\ref{eqDefU}) and~(\ref{eqDefUp}),
\begin{equation} \label{eqetabdef}
\eta_{b'} = 100 \times \frac{ ||\mathbf{ \tilde{b}  - \mathbf{b'}}||_{L^2} }{ ||\mathbf{b'}||_{L^2} }\,,
\end{equation}
where $||\cdot||_{L^2}$ is the $L^{2}$ norm \cite{Atkinson_1988}. For all the results shown in this article, we used vectors $\mathbf{b'}$ and $\mathbf{\tilde{b}}$ with 800 components [400 components each for $\bar U(\theta)$ and $\bar U'(\theta)$] in Eq.~(\ref{eqetabdef}), which we chose for a given protein shape so as to yield reference points with a uniform spacing in $\theta$ over the interval $0 \leq \theta \leq 2 \pi$. Figure~\ref{fig:4}(a) shows $\eta_{b'}$ in Eq.~(\ref{eqetabdef}) for the clover-leaf shapes in Eq.~(\ref{eqCdefclover}) with $s=3$ and various values of $\epsilon$. As expected, we find that $\eta_{b'}$ tends to decrease with increasing $N$ in Eq.~(\ref{equEigenSuperN}) with Eq.~(\ref{eqEigenFuncsN}), indicating that a greater accuracy of BVM solutions is obtained at larger $N$. The local minima of $\eta_{b'}$ in Fig.~\ref{fig:4}(a) correspond to values of $N$ that are multiples of $s$, which suggests that the accuracy of the BVM is improved if $N$ matches the protein symmetry. Figure~\ref{fig:4}(b) indicates that the convergence of BVM solutions with increasing $N$ can be improved substantially through an APD that allows for a nonuniform distribution of boundary points, which we discuss in Sec.~\ref{subsubsecAPD}.

We performed our BVM calculations in C++ using the arbitrary precision numerical library \textit{Arb} \cite{Johansson_arb_2017}. Unless stated otherwise, we allowed for sufficient numerical precision so that the boundary error $\eta_{b'} \leq 0.1 \%$ in Eq.~(\ref{eqetabdef}) and we obtained changes in $G$ and $\eta_{b'}$ of no more than $10^{-5}\%$ as the numerical precision was increased. We generated all figures in \textit{MATLAB} \cite{MATLAB_2021}. To speed up our calculations, we multi-threaded some of the source code of the \textit{Arb} library \cite{dagum_omp_1998}. Appendix~\ref{AppA} provides a more in-depth description of our computational implementation of the BVM, and discusses possible issues with the numerical solution of Eq.~(\ref{eqmeq}) arising from floating point errors and numerical instabilities. For the polygonal protein shapes considered here, with $P=5$ in Eq.~(\ref{eqDefCpoly}) with Eq.~(\ref{eqDefCpoly2}), and for clover-leaf protein shapes with large $s$ and/or $\epsilon$ in Eq.~(\ref{eqCdefclover}) we found it convenient to perform the BVM calculations with numerical precision greater than double precision (64 bits). In Appendix~\ref{AppB} we illustrate the extent to which double precision calculations could be used to approximate the BVM results described here.

In our BVM calculations we evaluate the bilayer thickness deformation energy $G$ in Eq.~(\ref{eqGArea}) by numerically computing Eq.~(\ref{eqGLine}) using the same 400 reference points employed to calculate $\eta_{b'}$ in Eq.~(\ref{eqetabdef}). To this end, we approximate $\bar G$ in Eq.~(\ref{eqGLine}) through Eq.~(\ref{equEigenSuperN}) with Eq.~(\ref{eqEigenFuncsN}),
\begin{equation} \label{eqGBVM}
\bar G \approx \frac{\mathbf{i}}{2}\int^{2\pi}_{ 0 } d \theta \bar l
\left( \bar U' - \bar U\mathbf{\hat{n}}\cdot \bar \nabla \right) \left( \bar f^{+}_{N} - \bar f^{-}_{N} \right) \big |_{\bar r=\bar C(\theta)}\,,
\end{equation}
where we have used Eq. (\ref{eqHelmholtz}) and the boundary conditions in Eqs.~(\ref{eqDefU}) and~(\ref{eqDefUp}). We have confirmed that, within the numerical accuracy used here, identical results for $\eta_{b'}$ and $\bar G$ are obtained with more than 400 reference points. Note that $\bar l$ in Eq.~(\ref{eqGBVM}) is real and that Eq.~(\ref{eqGBVM}) must evaluate to a real number, which means that the remaining terms in the integrand in Eq.~(\ref{eqGBVM}) evaluate to a purely imaginary number.

\begin{figure}[t!]{\includegraphics{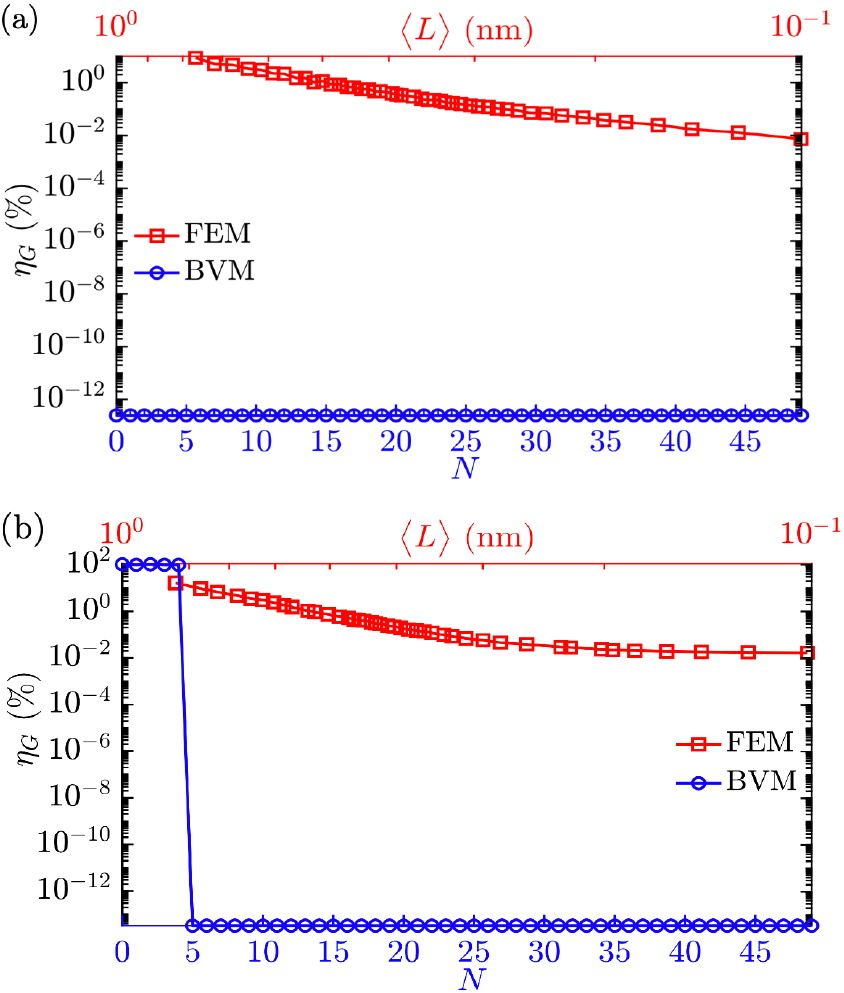}}
	\caption{Percentage difference between exact analytic and FEM (red curves) or BVM (blue curves) solutions for the bilayer thickness deformation energy, $\eta_{G}$ in Eq.~(\ref{eqetaGdef}), as a function of the average edge size $\langle L \rangle$ used in the FEM solution (upper axes) or the number of terms in Eq.~(\ref{equEigenSuperN}) with Eq.~(\ref{eqEigenFuncsN}) used in the BVM solution (lower axes) for (a) a cylindrical protein with $\bar R \lambda=2.3$~nm and $\bar U \lambda=0.3$~nm and (b) a crown-shaped protein with $\bar R \lambda=2.3$~nm, $\bar U_0 \lambda=-0.1$~nm, $\bar \beta \lambda=0.5$~nm, and $w=5$ in Eq.~(\ref{eqUBC}). We set $\bar U'=0$ for both panels.}
	\label{fig:5}
\end{figure}

We validated the BVM for bilayer thickness deformations against exact analytic solutions obtained for proteins with circular cross sections \cite{Huang_1986,Nielsen_Goulian_Andersen_1998,Wiggins_Phillips_2005,Haselwandter_Phillips_PLOS_2013,Haselwandter_Phillips_EPL_2013,Kahraman_Koch_Klug_Haselwandter_PRE_2016} and against FEM solutions \cite{Kahraman_Klug_Haselwandter_2014,Kahraman_Koch_Klug_Haselwandter_PRE_2016,Kahraman_Koch_Klug_Haselwandter_SR_2016} (see Fig.~\ref{fig:5}). In particular, we consider in Fig.~\ref{fig:5} cylindrical membrane proteins with constant $U$ and $U'$ [see Fig.~\ref{fig:5}(a)], for which the exact analytic solution of bilayer thickness deformations simply amounts to the zeroth-order terms in Eq.~(\ref{equEigenSuperN}) with Eq.~(\ref{eqEigenFuncsN}) \cite{Huang_1986,Nielsen_Goulian_Andersen_1998,Wiggins_Phillips_2005}. Furthermore, we consider in Fig.~\ref{fig:5} crown-shaped membrane proteins with circular cross section, constant $U'$, and the periodically varying $U(\theta)$ in Eq.~(\ref{eqUBC}) [see Fig.~\ref{fig:5}(b)], for which the exact analytic solution is obtained at order $N=w$ in Eq.~(\ref{equEigenSuperN}) with Eq.~(\ref{eqEigenFuncsN}) \cite{Haselwandter_Phillips_PLOS_2013,Haselwandter_Phillips_EPL_2013,Kahraman_Koch_Klug_Haselwandter_PRE_2016}. We quantified the level of agreement between the BVM and FEM solutions and the corresponding exact analytic solutions through the percentage difference in the calculated bilayer thickness deformation energy $G$ in Eq.~(\ref{eqGArea}),
\begin{equation} \label{eqetaGdef}
\eta_{G} = 100 \times \Biggr{|}\frac{G - G_\mathrm{analy} }{G_\mathrm{analy}}\Biggr{|}\,,
\end{equation}
where $G_\mathrm{analy}$ denotes the analytic solution \cite{Huang_1986,Nielsen_Goulian_Andersen_1998,Wiggins_Phillips_2005,Haselwandter_Phillips_PLOS_2013,Haselwandter_Phillips_EPL_2013,Kahraman_Koch_Klug_Haselwandter_PRE_2016} and $G$ denotes the corresponding BVM or FEM solutions. We found, as expected, excellent numerical agreement between the BVM and the aforementioned exact analytic solutions for $N=0$ [Fig.~\ref{fig:5}(a)] or $N \geq w$ [Fig.~\ref{fig:5}(b)] within floating point error. The FEM solutions in Fig.~\ref{fig:5} are, up to their expected numerical precision \cite{Kahraman_Koch_Klug_Haselwandter_PRE_2016}, in good agreement with the exact analytic and BVM solutions, with the agreement improving with decreasing average edge size $\langle L \rangle$ in the FEM grid. For both cylindrical and crown-shaped membrane proteins, we have $\eta_{G}\approx 0.01 \%$ for $\langle L\rangle \approx 0.1$~nm in the FEM solutions in Fig.~\ref{fig:5}.

\subsection{Nonuniform boundary point distributions}
\label{subsubsecAPD}

As illustrated in Figs.~\ref{fig:4} and~\ref{fig:5}, the BVM can provide a highly accurate method for calculating protein-induced bilayer thickness deformations. However, Fig.~\ref{fig:4}(a) also shows that, for large enough deviations from a circular protein cross section, accurate BVM solutions require a large number of terms in the Fourier-Bessel series in Eq.~(\ref{equEigenSuperN}) with Eq.~(\ref{eqEigenFuncsN}). The numerical performance of the BVM can be improved substantially, for non-circular protein cross sections, by choosing suitable, nonuniform boundary point distributions. In particular, we found that boundary point distributions that assign more points to, as viewed from the lipid bilayer, concave boundary regions yield a more rapid convergence of $G$ with increasing $N$. This can be understood intuitively by noting that, in the concave regions of a boundary curve, different sections of the boundary curve can produce overlapping bilayer thickness deformation fields, inducing protein self-interactions. One expects that higher-order terms in the Fourier-Bessel series in Eq.~(\ref{equEigenSuperN}) with Eq.~(\ref{eqEigenFuncsN}) are required to capture such interactions \cite{Haselwandter_Phillips_EPL_2013,Kahraman_Koch_Klug_Haselwandter_PRE_2016}.

\begin{figure}[t!]{\includegraphics{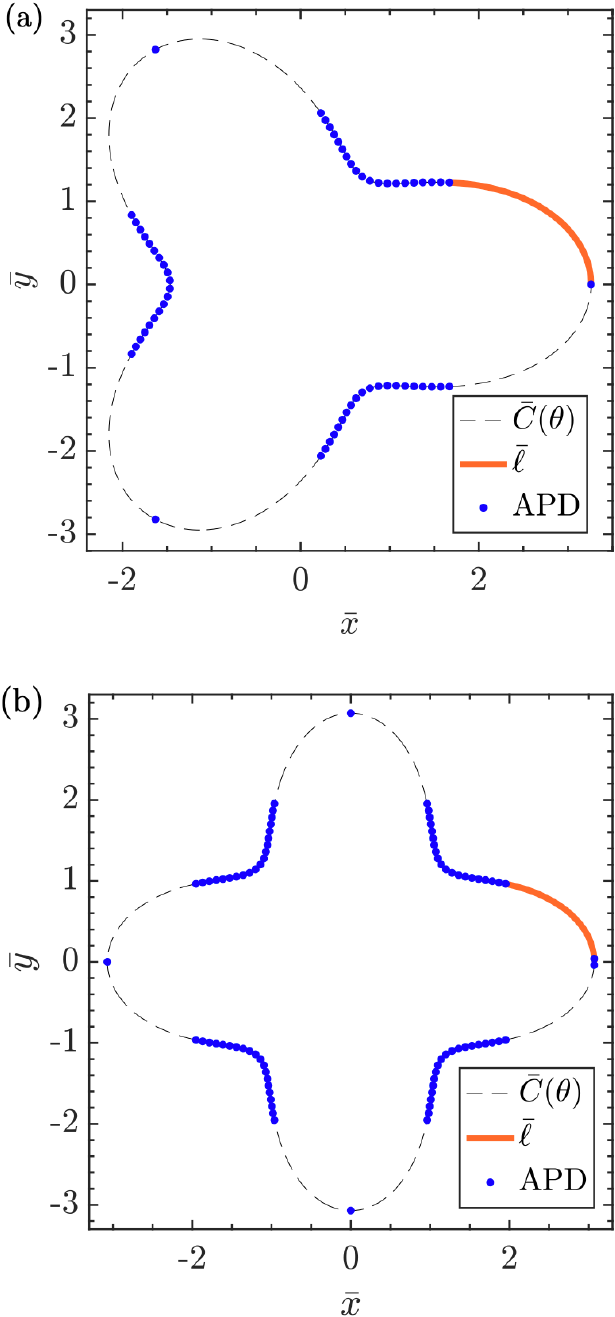}}
	\caption{Illustration of the APD method used to increase the numerical efficiency of BVM solutions for (a) a three-fold clover-leaf protein shape ($s=3$) and (b) a four-fold clover-leaf protein shape ($s=4$) in Eq.~(\ref{eqCdefclover}). The boundary points used for the BVM solutions are indicated by blue dots. In panel (a) we set $\epsilon=0.38$, $N=31$, and $\Omega=0.62$ for the gap length $\bar \ell$ in Eq.~(\ref{eqelldef}). In panel (b) we set $\epsilon=0.30$, $N=42$, and $\Omega=0.72$. For both panels we set $\bar R \lambda\approx 2.3$~nm. To achieve an approximately periodic distribution of boundary points for even $s$, we duplicated in panel (b) the boundary point at the right-most apex, and slightly offset the resulting two boundary points along the bilayer-protein interface (see main text). The values of $N$ (numbers of boundary points) considered here were chosen for illustrative purposes. We generally employ values of $N$ greater than those considered here so as to meet the numerical precision criteria imposed here (see main text).}
	\label{fig:6}
\end{figure}

To assign more boundary points to the concave boundary regions of clover-leaf and (finite-$P$) polygonal protein shapes, we employ an APD of the BVM boundary points. In the APD method, we distribute the $2N+1$ boundary points such that boundary points are always assigned to the apex points along the bilayer-protein boundary curves furthest away from the protein center (see Fig.~\ref{fig:6}). We distribute the remaining boundary points along the sections of the bilayer-protein boundary curves that are an arc length $\bar l\geq \bar \ell$ away from the apex points, such that these points are uniformly spaced in arc length, with the gap length 
\begin{equation} \label{eqelldef}
\bar \ell = \Omega \frac{\bar \Gamma}{2s} \,,
\end{equation}
where the gap factor $\Omega$ satisfies $0 < \Omega < 1$, $\bar \Gamma$ is the (dimensionless) protein circumference, and $s$ is the symmetry of the clover-leaf or polygonal protein shape [see also Eqs.~(\ref{eqCdefclover}) and~(\ref{eqDefCpoly})] (Fig.~\ref{fig:6}). For even values of $s$, we consider in our APD method the general solution in Eq.~(\ref{equEigenSuperN}) with Eq.~(\ref{eqEigenFuncsN}) for $N = s M/2$ with integers $M \geq 3$. For odd values of $s$, we allow in Eq.~(\ref{equEigenSuperN}) with Eq.~(\ref{eqEigenFuncsN}) for $N = s M/2$ for even integers $M \geq 3$, and for $N = \left(s M-1\right)/2$ for odd integers $M \geq 3$. To achieve an approximately periodic distribution of boundary points for even $s$ and for odd $s$ with even $M$, we found it convenient to duplicate one of the apex boundary points, with a slight offset in the duplicated boundary points by an equal arc length distance from the apex [see Fig.~\ref{fig:6}(b)]. For greater numerical accuracy, this distance from the apex could be optimized so as to reduce the boundary error $\eta_{b'}$ in Eq.~(\ref{eqetabdef}), but we found it sufficient here to set it equal to one-half the arc length spacing between the boundary points in the concave boundary regions. Unless stated otherwise, we used the APD method for all BVM calculations described in this article, fixing $\Omega$ in Eq.~(\ref{eqelldef}) and $N$ in Eq.~(\ref{equEigenSuperN}) with Eq.~(\ref{eqEigenFuncsN}) such that the boundary error $\eta_{b'} \leq 0.1 \%$ in Eq.~(\ref{eqetabdef}) and we obtained changes in $G$ and $\eta_{b'}$ of no more than $10^{-5}\%$ as the numerical precision was increased.

As illustrated in Fig.~\ref{fig:4}(b) for clover-leaf shapes, the APD method employed here improves considerably the convergence of the BVM with increasing $N$, particularly for proteins that show substantial deviations from a circular cross section. As a result, a given numerical accuracy of BVM solutions can be achieved with smaller $N$. We note that, for proteins with (discrete) rotational symmetry, the Fourier-Bessel series in Eq.~(\ref{equEigenSuperN}) with Eq.~(\ref{eqEigenFuncsN}) must show the same symmetry. Indeed, in our BVM calculations we find that, within the numerical precision employed here, the coefficients of terms in Eq.~(\ref{equEigenSuperN}) with Eq.~(\ref{eqEigenFuncsN}) that break the protein symmetry take values equal to zero. On this basis one could, for a given protein symmetry, further improve the numerical efficiency of the BVM by using the protein symmetry to remove some of the terms in the Fourier-Bessel series in Eq.~(\ref{equEigenSuperN}) with Eq.~(\ref{eqEigenFuncsN}). For the scenarios considered here the BVM was efficient enough so as not to require such further refinement.

\begin{figure}[t!]{\includegraphics{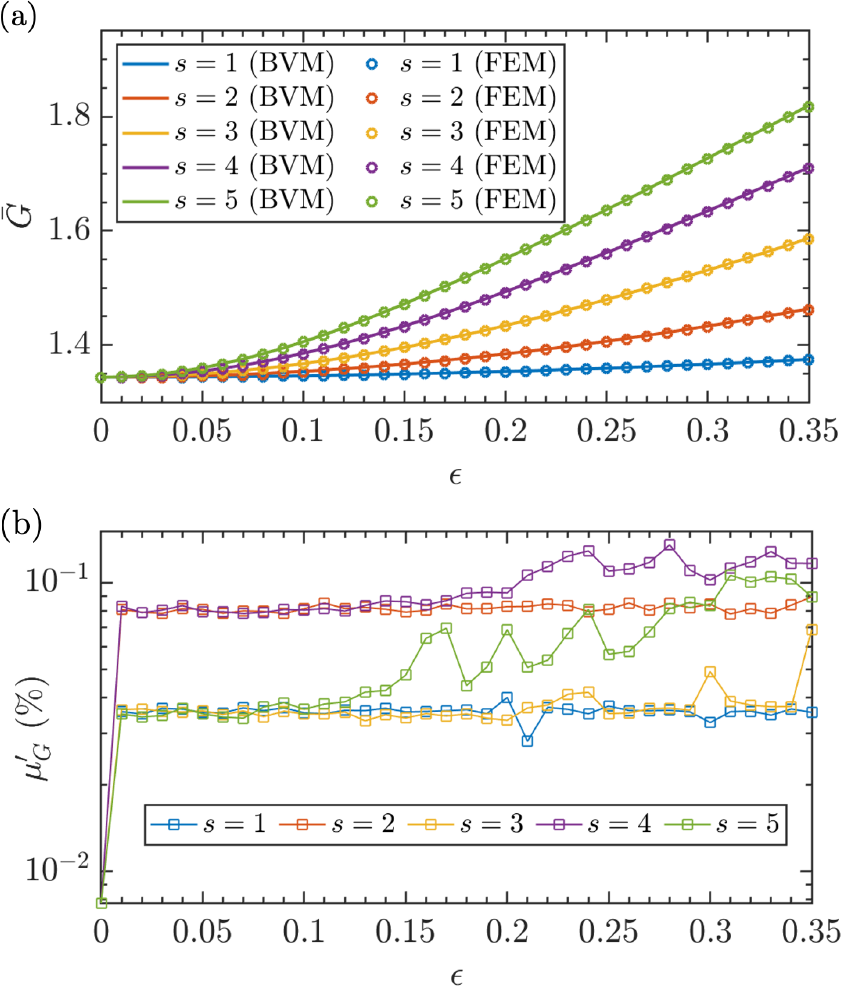}}
	\caption{Comparing BVM and FEM solutions for the elastic energy of protein-induced bilayer thickness deformations. (a) Bilayer thickness deformation energy, $\bar G$ in Eq.~(\ref{eqGAreaDL}), obtained using BVM and FEM solutions for $\bar u$ in Eq.~(\ref{equdef}) and (b) corresponding percentage difference between the BVM and FEM solutions for $\bar G$, $\mu_{G}'$ in Eq.~(\ref{eqmuGdef}), for the clover-leaf protein shapes in Eq.~(\ref{eqCdefclover}) as a function of $\epsilon$ with the indicated values of $s$, $\bar R\lambda\approx2.3$~nm, $\bar U \lambda=0.3$~nm, and $\bar U'=0$. For the FEM solutions we employed an average edge size $\langle L \rangle \approx 0.1$ nm.}
	\label{fig:7}
\end{figure}

Figure~\ref{fig:7} illustrates the calculation of the bilayer thickness deformation energy, $G$ in Eq.~(\ref{eqGArea}), using the BVM with APD for the clover-leaf protein shapes in Eq.~(\ref{eqCdefclover}) with various protein symmetries $s$ and deviations from a circular cross section, $\epsilon$. As expected \cite{Haselwandter_Phillips_PLOS_2013,Kahraman_Klug_Haselwandter_2014}, we find in Fig.~\ref{fig:7}(a) that $G$ increases with increasing $s$ and $\epsilon$. We also show in Fig.~\ref{fig:7} the corresponding results obtained from the FEM with an average edge size $\langle L \rangle\approx 0.1$~nm. In Fig.~\ref{fig:7}(b) we quantify the agreement between our BVM and FEM results through the percentage difference in $G$,
\begin{equation}  \label{eqmuGdef}
\mu_{G}' = 100 \times \Biggr{|}\frac{G_\mathrm{BVM} - G_\mathrm{FEM}}{G_\mathrm{FEM} }\Biggr{|}\,,
\end{equation}
where $G_\mathrm{BVM}$ and $G_\mathrm{FEM}$ correspond to the values of $G$ in Eq.~(\ref{eqGArea}) obtained through the BVM and the FEM \cite{Kahraman_Klug_Haselwandter_2014,Kahraman_Koch_Klug_Haselwandter_PRE_2016,Kahraman_Koch_Klug_Haselwandter_SR_2016}, respectively. We find in Fig.~\ref{fig:7} that the BVM and FEM solutions yield excellent agreement for the energy of protein-induced bilayer thickness deformations for non-circular as well as circular protein cross sections, with the level of agreement between BVM and FEM solutions being in line with the accuracy of the FEM solutions expected from Fig.~\ref{fig:5}.

\section{Analytic approximation of the bilayer thickness deformation energy}
\label{secAnalytic}

For membrane inclusions with circular cross section, the solution for the thickness deformation field $\bar u(\bar r,\theta)$ in Eq.~(\ref{equEigenSuper}) with Eq.~(\ref{eqEigenFuncs}) and the bilayer thickness deformation energy in Eq.~(\ref{eqGLine}) yield exact analytic expressions for the energy of protein-induced bilayer thickness deformations for arbitrary (angular) variations in the bilayer-protein boundary conditions \cite{Huang_1986,Dan_Pincus_Safran_1993,Dan_Berman_Pincus_Safran_1994, Haselwandter_Phillips_PLOS_2013,Haselwandter_Phillips_EPL_2013,Kahraman_Klug_Haselwandter_2014,Kahraman_Koch_Klug_Haselwandter_PRE_2016}. The purpose of this section is to develop, on this basis, a simple analytic scheme for estimating the energy of protein-induced bilayer thickness deformations for membrane proteins with non-circular cross sections. In Sec.~\ref{secDependShape} we show that, for many protein shapes, these simple analytic estimates agree remarkably well with the corresponding BVM solutions.

For a single membrane inclusion with circular cross section and arbitrary (angular) variations in $U(\theta)$ and $U'(\theta)$, the exact solution of the Euler-Lagrange equation in Eq.~(\ref{eqHelmholtz}) is given by Eq.~(\ref{equEigenSuper}) with Eq.~(\ref{eqEigenFuncs}), and the corresponding bilayer thickness deformation energy follows from Eq.~(\ref{eqGLine}) \cite{Haselwandter_Phillips_PLOS_2013,Haselwandter_Phillips_EPL_2013,Kahraman_Klug_Haselwandter_2014,Kahraman_Koch_Klug_Haselwandter_PRE_2016}. For the choices for $U(\theta)$ and $U'(\theta)$ in Eqs.~(\ref{eqUBC}) and~(\ref{eqUpBC}), one thus finds the bilayer thickness deformation energy
\begin{equation}\label{eqGanalytic}
		\begin{split}
			\bar G_\mathrm{analy} = \pi \bar R_\mathrm{analy} &(\bar \nu_+ - \bar \nu_-) \bigg{[} \bar U^2_0 \bar E_0 + \bar U'^2_0 \bar F_0 + \bar U_0 \bar U'_0 \bar H_0 \\
			 + \frac{1}{2} \bigg( \bar \beta^2 & \bar E_w + \bar \gamma^2 \bar F_v + \delta_{wv} \bar \beta \, \bar \gamma \bar H_w \bigg) \bigg{]}\bigg{\rvert}_{\bar r = \bar R_{\mathrm{analy}}}
 		\end{split}
\end{equation}
with $v > 0$ and $w > 0$, where $\bar R_\mathrm{analy}$ is the radius of the circular protein cross section, $\delta_{wv}$ is the Kronecker delta, and we have defined
\begin{equation}\label{eqDEFH}
	\begin{split}
		\bar D_q &= K_q(\sqrt{\bar \nu_+} \bar r) \partial_{\bar r} K_q(\sqrt{\bar \nu_-} \bar r) - K_q(\sqrt{\bar \nu_-} \bar r) \partial_{\bar r} K_q(\sqrt{\bar \nu_+} \bar r)\,,\\
		\bar E_q &= \frac{\left[\partial_{\bar r} K_q(\sqrt{\bar \nu_+}\bar r) \right] \left[ \partial_{\bar r} K_q(\sqrt{\bar \nu_-} \bar r)\right]}{\bar D_q}\,,\\
		\bar F_q &= \frac{K_q(\sqrt{\bar \nu_+} \bar r) K_q(\sqrt{\bar \nu_-} \bar r)}{\bar D_q}\,,\\
		\bar H_q &= \frac{K_q(\sqrt{\bar \nu_+} \bar r) \partial_{\bar r} K_q(\sqrt{\bar \nu_-} \bar r) + K_q(\sqrt{\bar \nu_-} \bar r) \partial_{\bar r} K_q(\sqrt{\bar \nu_+} \bar r)}{\bar D_q}\,,
	\end{split}
\end{equation}
where $q=0,1,\dots$, $K_q$ denotes the $q^\mathrm{th}$ order modified Bessel function of the second kind, and $\partial_{\bar r}$ denotes the partial derivative with respect to $\bar r$. The zeroth order terms in Eq.~(\ref{eqGanalytic}) are the contributions to $\bar G_\mathrm{analy}$ due to the constant $\bar U_0$ and $\bar U_0^\prime$ in Eqs.~(\ref{eqUBC}) and~(\ref{eqUpBC}), while the remaining terms encapsulate the effects of the variations in $U(\theta)$ and $U'(\theta)$ in Eqs.~(\ref{eqUBC}) and~(\ref{eqUpBC}) on $\bar G_\mathrm{analy}$. We use here Eq.~(\ref{eqGanalytic}) to analytically estimate the bilayer thickness deformation energy of membrane proteins with non-circular cross sections. To this end, we choose $\bar R_\mathrm{analy}$ in Eq.~(\ref{eqGanalytic}) such that the circumference of the circular membrane inclusion considered in Eq.~(\ref{eqGanalytic}) is equal to the circumference of the membrane protein under consideration,
\begin{equation} \label{Ranaly}
\bar R_\mathrm{analy} = \frac{\bar \Gamma}{2 \pi}\,,
\end{equation}
where, for the clover-leaf and polygonal boundary curves in Eqs.~(\ref{eqCdefclover}) and~(\ref{eqDefCpoly}) with Eq.~(\ref{eqDefCpoly2}), the protein circumference $\Gamma$ follows from $\bar \Gamma= \int^{2\pi}_{0} d\theta \bar l$, where, as in Eq.~(\ref{eqGLine}), $\bar l$ is the (dimensionless) line element.

\begin{figure}[t!]{\includegraphics{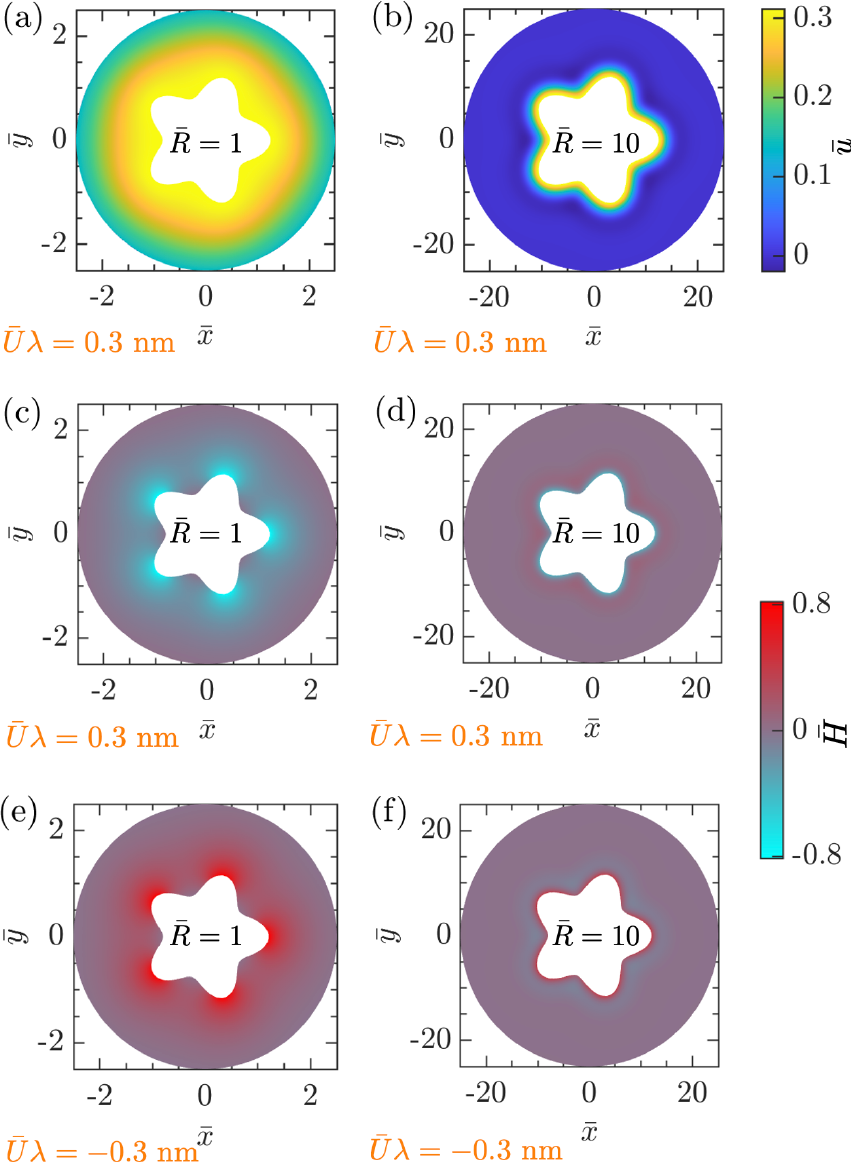}}
	\caption{Color maps of the bilayer thickness deformation footprints due to clover-leaf protein shapes with (a) $\bar R=1$ and (b) $\bar R=10$ in Eq.~(\ref{eqCdefclover}) for $s=5$, $\epsilon=0.2$, $\bar U \lambda=0.3$~nm in Eq.~(\ref{eqUBC}), and $\bar U'=0$. Panels (c) and (d) show the mean curvature in units of $1/\lambda$, $\bar H = \lambda H$, associated with the thickness deformation fields in panels (a) and (b), respectively, while panels (e) and (f) show the corresponding mean curvature maps obtained for $\bar U \lambda=-0.3$~nm in Eq.~(\ref{eqUBC}) rather than $\bar U \lambda=0.3$~nm. We set $2\bar a \lambda=3.2$~nm for all panels. All results were obtained through the BVM.}
	\label{fig:8}
\end{figure}

\begin{figure}[t!]{\includegraphics{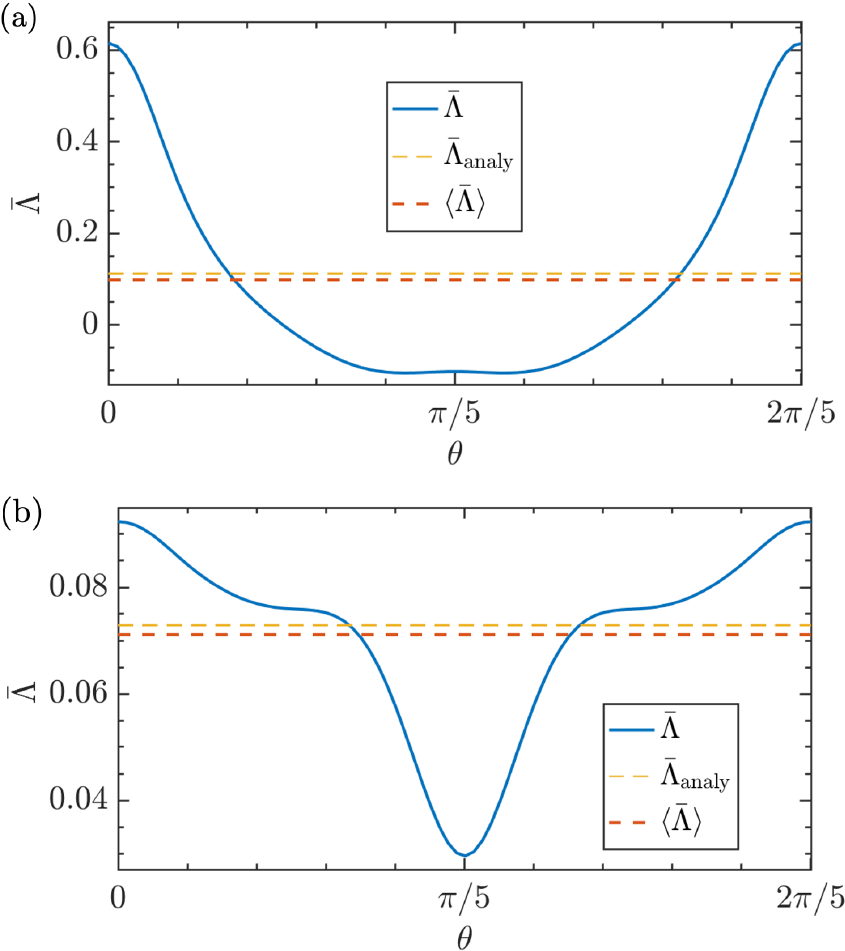}}
	\caption{Line tension along the bilayer-protein boundary, $\bar \Lambda$ in Eq.~(\ref{eqdefLambda}), as a function of $\theta$ for (a) the protein shape in Fig.~\ref{fig:8}(a) and (b) the protein shape in Fig.~\ref{fig:8}(b), calculated using the same parameter values as in Fig.~\ref{fig:8}. The red dashed lines show the average of $\bar \Lambda(\theta)$ in Eq.~(\ref{eqdefLambda}) over the interval $0 \leq \theta \leq 2 \pi/5$, $\langle \bar \Lambda \rangle$. The yellow dashed lines show $\bar \Lambda_\mathrm{analy}=\bar G_\mathrm{analy}/\bar \Gamma$, where $\bar G_\mathrm{analy}$ is given by Eq.~(\ref{eqGanalytic}) and $\bar \Gamma$ is the protein circumference in Eq.~(\ref{Ranaly}).}
	\label{fig:9}
\end{figure}

The analytic estimate of the thickness deformation energy in Eq.~(\ref{eqGanalytic}) captures, for the choice of $\bar R_\mathrm{analy}$ in Eq.~(\ref{Ranaly}), effects related to the overall shape of membrane proteins. However, Eq.~(\ref{eqGanalytic}) does not capture effects due to strong local variations in the protein cross section. For instance, the clover-leaf shapes in Eq.~(\ref{eqCdefclover}) can give, for large enough $\epsilon$ and $s$, protein cross sections with pronounced invaginations. If the protein size $R$ is comparable to the decay length of bilayer thickness deformations, $\lambda$ in Eq.~(\ref{eqDeflambda}), such protein invaginations can yield overlaps in the protein-induced lipid bilayer thickness deformations due to different portions of the bilayer-protein interface, resulting in protein self-interactions [see Fig.~\ref{fig:8}(a)]. As $\bar R$ is increased, these overlaps in protein-induced bilayer thickness deformations become less pronounced [see Fig.~\ref{fig:8}(b)]. Depending on the value of $\bar R$, one thus obtains distinct distributions of the mean curvature of $\bar u$ about the protein [see Figs.~\ref{fig:8}(c,d)], which also depend on the value and sign of $\bar U$ [see Figs.~\ref{fig:8}(e,f)].

To quantify the protein self-interactions suggested by Fig.~\ref{fig:8} it is useful to define, based on Eq.~(\ref{eqGLine}), the line tension along the bilayer-protein interface,
\begin{equation} \label{eqdefLambda}
\bar \Lambda \equiv \left[\bar U'(\theta) \bar \nabla^{2} \bar u - \bar U(\theta) \mathbf{\hat{n}} \cdot \bar \nabla^{3} \bar u \right] \big |_{\bar r=\bar C(\theta)}\,,
\end{equation}
where we used Eqs.~(\ref{eqDefU}) and~(\ref{eqDefUp}). In Figs.~\ref{fig:9}(a) and~\ref{fig:9}(b) we compare, for the protein shapes in Figs.~\ref{fig:8}(a) and~\ref{fig:8}(b) with constant $\bar U > 0$ and $\bar U^\prime = 0$, the line tensions $\bar \Lambda$ in Eq.~(\ref{eqdefLambda}) and their average values $\langle\bar\Lambda\rangle$ to the corresponding (constant) $\bar \Lambda$ associated with $\bar G_\mathrm{analy}$ in Eq.~(\ref{eqGanalytic}), which we denote by $\bar \Lambda_{\mathrm{analy}}$. As expected, Fig.~\ref{fig:9} shows that the variations in $\bar \Lambda$ are more pronounced for smaller clover-leaf protein shapes. We also find in Fig.~\ref{fig:9} that $\langle\bar\Lambda\rangle < \bar \Lambda_{\mathrm{analy}}$, with a larger $\big|\langle\bar\Lambda\rangle - \bar \Lambda_{\mathrm{analy}}\big|$ for smaller $\bar R$ in Fig.~\ref{fig:9}.

Interestingly, we can have $\bar \Lambda < 0$ in Fig.~\ref{fig:9}(a) for the smaller clover-leaf protein shape in Fig.~\ref{fig:8}(a), while $\bar \Lambda > 0$ in Fig.~\ref{fig:9}(b) for the larger clover-leaf protein shape in Fig.~\ref{fig:8}(b). The regime with $\bar \Lambda < 0$ in Fig.~\ref{fig:9}(a) can be understood by noting that, with a constant $\bar U > 0$ and $\bar U^\prime = 0$, $\bar \Lambda$ in Eq.~(\ref{eqdefLambda}) is directly proportional to the change in the mean curvature of $\bar u$ at the protein boundary, in the direction perpendicular to the protein-bilayer boundary and into the bilayer $(-\mathbf{\hat{n}})$. For the points along the clover-leaf boundary closest and furthest away from the protein center, $\mathbf{\hat{n}}$ is anti-parallel with the radial direction $\mathbf{\hat{r}}$. For the points along the clover-leaf boundary shape in Figs.~\ref{fig:8}(a) and~\ref{fig:8}(b) furthest away from the protein center the mean curvature is negative with the sign convention used here and decreases in magnitude as one radially moves away from the protein boundary, yielding $\bar \Lambda > 0$ [Figs.~\ref{fig:8}(c,d)]. In contrast, for the points along the clover-leaf boundary shape closest to the protein center in Fig.~\ref{fig:8}(a) [but not Fig.~\ref{fig:8}(b)], the mean curvature is approximately zero at the protein boundary and decreases as one radially moves away from the protein boundary, yielding $\bar \Lambda < 0$. With a different sign convention for the mean curvature or a protein with a constant $\bar U<0$ rather than $\bar U>0$, analogous considerations apply [Figs.~\ref{fig:8}(e,f)]. Thus, protein self-interactions can effectively lower the energy cost of protein-induced lipid bilayer thickness deformations, in analogy to the energetically favorable bilayer-thickness-mediated protein interactions found for identical membrane proteins in close enough proximity \cite{Dan_Pincus_Safran_1993,Dan_Berman_Pincus_Safran_1994,Ursell2007,harroun99,goforth03,botelho06,Phillips_Ursell_Wiggins_Sens_2009,grage11,Haselwandter_Phillips_EPL_2013,Kahraman_Klug_Haselwandter_2014,Haselwandter_Wingreen_2014,milovanovic15,Kahraman_Koch_Klug_Haselwandter_PRE_2016,Kahraman_Koch_Klug_Haselwandter_SR_2016,pollard18}.

\section{Dependence of bilayer thickness deformation energy on protein shape}
\label{secDependShape}

In this section we survey the dependence of the bilayer thickness deformation energy in Eq.~(\ref{eqGArea}) on the shape of membrane proteins. In particular, we allow for three distinct, not mutually exclusive, modes for breaking rotational symmetry about the protein center (see also Sec.~\ref{subsecProteinShape}). In Sec.~\ref{secShapeConst} we take the bilayer-protein boundary conditions to be constant along the protein circumference, but allow for protein cross sections that break rotational symmetry about the protein center. In Sec.~\ref{secShapeU} we explore the effect of variations in the protein hydrophobic thickness on protein-induced bilayer thickness deformations. Finally, in Sec.~\ref{secShapeUp} we study protein-induced bilayer thickness deformations for proteins that show variations in the bilayer-protein contact slope along the bilayer-protein boundary. To test the analytic approximation of the bilayer thickness deformation energy described in Sec.~\ref{secAnalytic} we compare, for all three scenarios considered in Secs.~\ref{secShapeConst}--\ref{secShapeUp}, our BVM results to the corresponding analytic estimates by computing the signed percent error 
\begin{equation} \label{eqxiG}
\xi_{G} = 100\times\frac{\bar G_{\mathrm{analy}} - \bar G}{\bar G}\,,
\end{equation}
where $\bar G$ is the thickness deformation energy in Eq.~(\ref{eqGAreaDL}) obtained through the BVM and the corresponding analytic estimate $\bar G_{\mathrm{analy}}$ is given by Eq.~(\ref{eqGanalytic}) with Eq.~(\ref{eqDEFH}).

\subsection{Constant bilayer-protein boundary conditions}
\label{secShapeConst}

\begin{figure}[t!]{\includegraphics{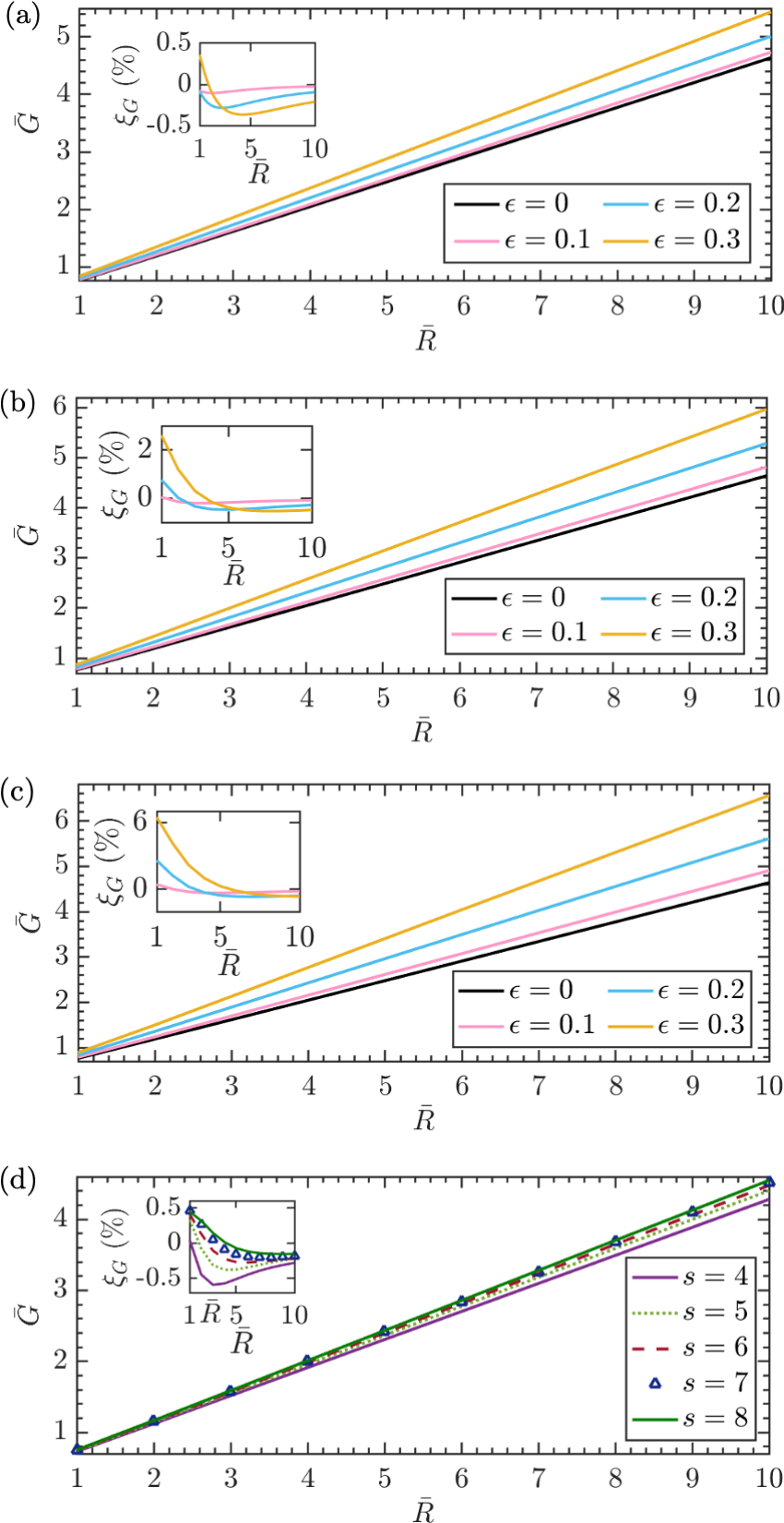}}
	\caption{Bilayer thickness deformation energy $\bar G$ in Eq.~(\ref{eqGAreaDL}) calculated using the BVM (see Sec.~\ref{secBVM}) as a function of protein size $\bar R$ for clover-leaf protein shapes with (a) $s=3$, (b) $s=4$, and (c) $s=5$ in Eq.~(\ref{eqCdefclover}) with the indicated values of $\epsilon$, and (d) polygonal protein shapes with the indicated values of $s$ and $P=5$ in Eq.~(\ref{eqDefCpoly}). For all panels we set $\bar U \lambda=0.3$~nm and $\bar U'=0$. The insets show the signed percent error $\xi_{G}$ in Eq.~(\ref{eqxiG}) for the corresponding analytic approximations $\bar G_{\text{analy}}$ in Eq.~(\ref{eqGanalytic}).
	}
	\label{fig:10}
\end{figure}

\begin{figure}[t!]{\includegraphics{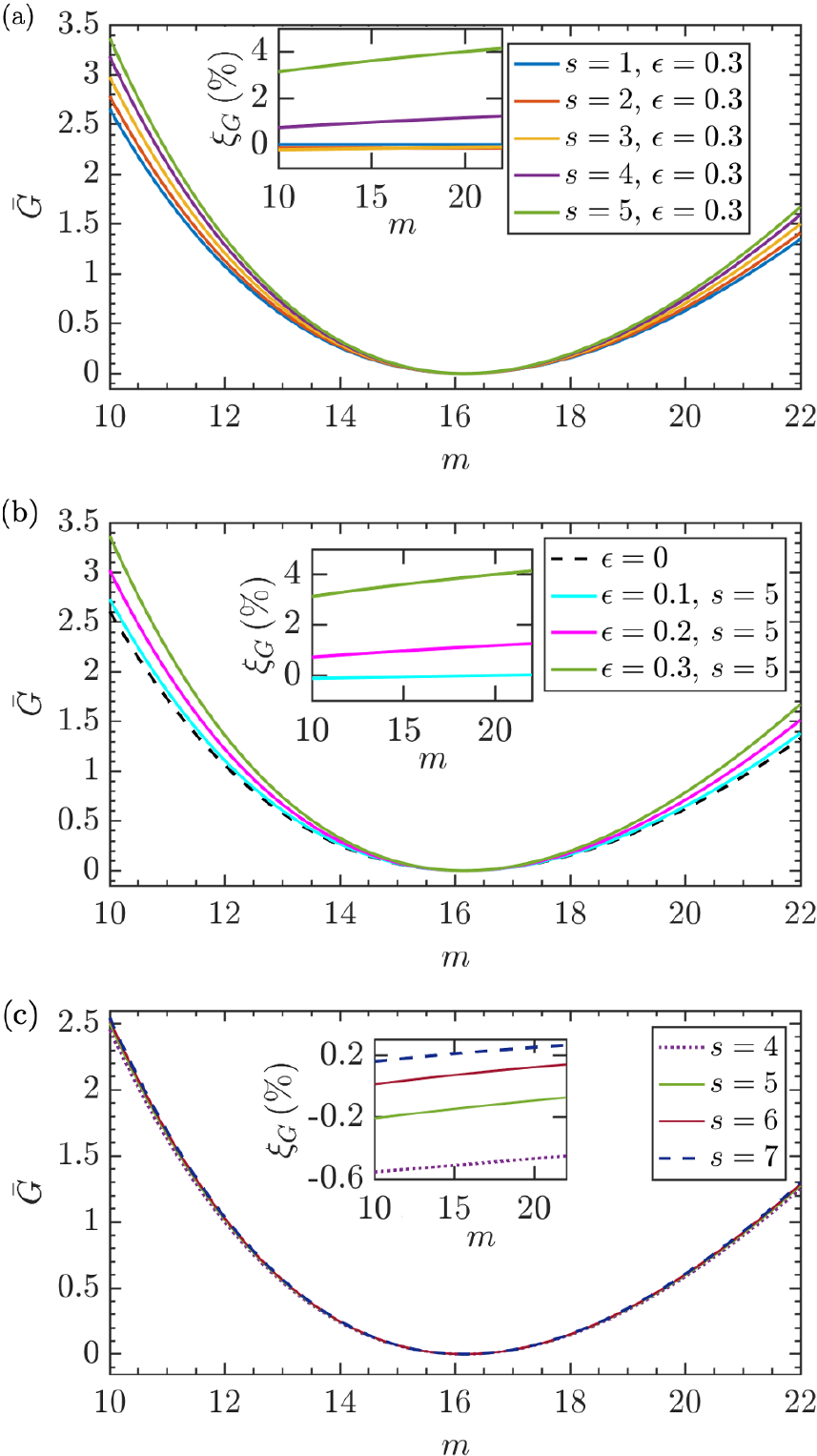}}
	\caption{Bilayer thickness deformation energy $\bar G$ in Eq.~(\ref{eqGAreaDL}) calculated using the BVM (see Sec.~\ref{secBVM}) as a function of lipid chain length $m$ in Eq.~(\ref{eqDefm}) for (a) clover-leaf protein shapes with $\epsilon = 0.3$ and the indicated values of $s$, (b) clover-leaf protein shapes with $s=5$ and the indicated values of $\epsilon$, and (c) polygonal protein shapes with the indicated values of $s$ and $P=5$ in Eq.~(\ref{eqDefCpoly}). For all panels we set $\bar U' = 0$ and $\bar W \lambda=3.8$~nm in Eq.~(\ref{eqUWa}), and $\bar R \lambda\approx2.3$~nm. The insets show the signed percent error $\xi_{G}$ in Eq.~(\ref{eqxiG}) for the corresponding analytic approximations $\bar G_{\text{analy}}$ in Eq.~(\ref{eqGanalytic}). We always have $\left|\bar U \right| > 0$ for the $m$-discretization used here.
	}
	\label{fig:11}
\end{figure}

In Fig.~\ref{fig:10} we consider the energy of protein-induced bilayer thickness deformations for clover-leaf [see Figs.~\ref{fig:10}(a,b,c)] and polygonal [see Fig.~\ref{fig:10}(d)] protein shapes as a function of protein size $\bar R$ with a constant $\bar U \neq 0$ and $\bar U'=0$. Previous work on the lipid bilayer thickness deformations induced by proteins with circular cross section \cite{Wiggins_Phillips_2004,Wiggins_Phillips_2005} suggests that, for $\bar R \gg 1$, $\bar G$ increases approximately linearly with $\bar R$. We find in Fig.~\ref{fig:10} that we also approximately have $\bar G \propto \bar R$ for non-circular protein cross sections, with the (positive) constant of proportionality depending on the protein shape. The analytic estimates $\bar G_{\mathrm{analy}}$ obtained from Eq.~(\ref{eqGanalytic}) match $\bar G$ in Fig.~\ref{fig:10} within approximately $10\%$, with particularly small magnitudes of the signed percent error $\xi_{\bar G}$ for the polygonal protein shapes in Fig.~\ref{fig:10}(d). Note that for protein sizes $R$ comparable to the decay length $\lambda$ we generally have $\xi_{\bar G} > 0$ in Fig.~\ref{fig:10}, indicating that protein self-interactions tend to lower the energy cost of protein-induced bilayer thickness deformations in Fig.~\ref{fig:10}.

The energy cost of protein-induced bilayer thickness deformations depends crucially on the unperturbed lipid bilayer thickness, which can be varied by changing the lipid chain length $m$ in Eq.~(\ref{eqDefm}) \cite{Rawicz_Olbrich_McIntosh_Needham_Evans_2000,Perozo_Kloda_Cortes_Martinac_2002,Andersen_Koeppe_2007,Phillips_Ursell_Wiggins_Sens_2009}. In Fig.~\ref{fig:11} we plot $\bar G$ for clover-leaf [see Figs.~\ref{fig:11}(a,b)] and polygonal [see Fig.~\ref{fig:11}(c)] protein shapes as a function of the lipid chain length $m$ with $\bar U'=0$. We used a protein hydrophobic thickness  $\bar W \lambda= 3.8$~nm, which matches the unperturbed lipid bilayer thickness for $m\approx16$. In Fig.~\ref{fig:11}(a) we consider clover-leaf protein shapes with different symmetries $s$ and the same value of $\epsilon$, while in Fig.~\ref{fig:11}(b) we consider clover-leaf protein shapes with different values of $\epsilon$ and the same symmetry $s$. Similarly as in Fig.~\ref{fig:10} we find that deviations from a circular protein cross section increase $\bar G$. Furthermore, similarly as in Fig.~\ref{fig:10}, the dependence of $\bar G$ on $m$ in Fig.~\ref{fig:11} is very well captured by the analytic approximation $\bar G_{\mathrm{analy}}$ in Eq.~(\ref{eqGanalytic}), suggesting that the increase in $\bar G$ for clover-leaf and polygonal protein cross sections compared to circular protein cross sections results primarily from the increase in the length of the bilayer-protein boundary $\bar \Gamma$ due to deviations from a circular protein cross section.

\begin{figure}[t!]{\includegraphics{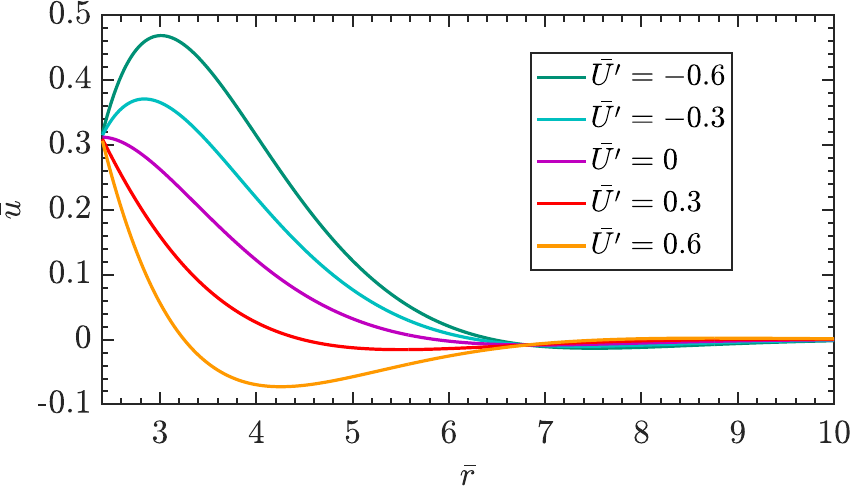}}
	\caption{Bilayer thickness deformation profile $\bar u$ due to a protein with a circular cross section as a function of the radial distance from the protein center, $\bar r=r/\lambda$, obtained from the exact analytic solution in Eq.~(\ref{equEigenSuper}) with Eq.~(\ref{eqEigenFuncs}) for the indicated values of $\bar U^\prime$. We set $\bar U \lambda=0.3$~nm and $\bar R \lambda=2.3$~nm.}
	\label{fig:12}
\end{figure}

\begin{figure}[t!]{\includegraphics{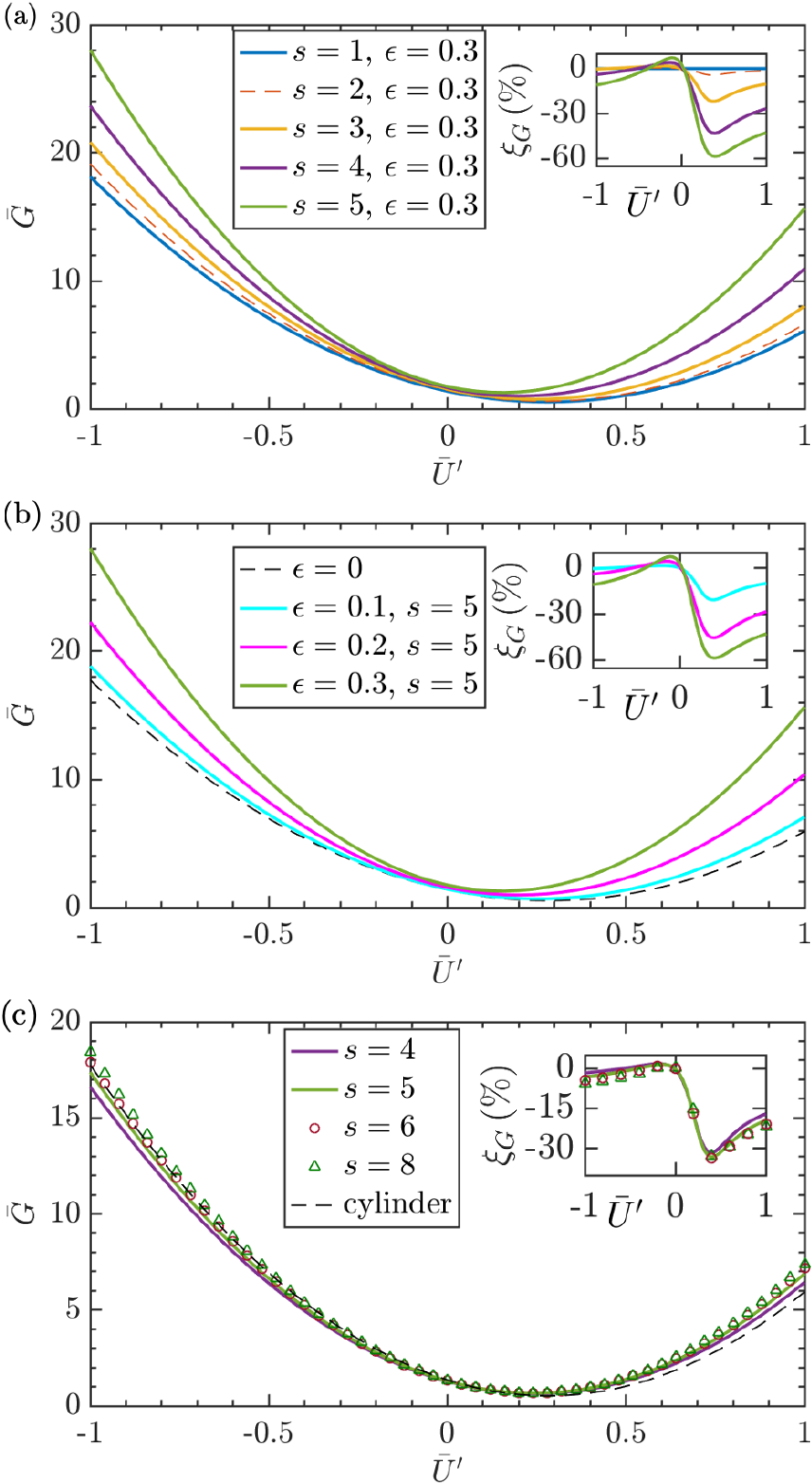}}
	\caption{Bilayer thickness deformation energy $\bar G$ in Eq.~(\ref{eqGAreaDL}) calculated using the BVM (see Sec.~\ref{secBVM}) as a function of the bilayer-protein contact slope $\bar{U'}$ for (a) clover-leaf protein shapes with $\epsilon = 0.3$ and the indicated values of $s$, (b) clover-leaf protein shapes with $s=5$ and the indicated values of $\epsilon$, and (c) polygonal protein shapes with the indicated values of $s$ and $P=5$, and cylindrical protein shapes with a circular cross section of radius $\bar R$. For all panels we set $\bar R \lambda=2.3$~nm and $\bar U \lambda=0.3$~nm. The insets show the signed percent error $\xi_{G}$ in Eq.~(\ref{eqxiG}) for the corresponding analytic approximations $\bar G_{\text{analy}}$ in Eq.~(\ref{eqGanalytic}).
	}
	\label{fig:13}
\end{figure}

In Figs.~\ref{fig:10} and~\ref{fig:11} we set, in line with previous work on gramicidin channels and MscL \cite{Huang_1986,Wiggins_Phillips_2004,Wiggins_Phillips_2005}, $U^\prime = 0$. However, the most suitable choice for the boundary conditions on the gradient of $u$ at the bilayer-protein interface has been a matter of debate \cite{Nielsen_Goulian_Andersen_1998,Nielsen_Andersen_2000,Argudo_Bethel_Marcoline_Wolgemuth_Grabe_2017}, and may depend on the particular membrane protein and lipid species under consideration. In particular, $U^\prime$ may differ from zero or vary along the bilayer-protein interface, or the gradient of $u$ at the bilayer-protein interface may satisfy natural boundary conditions with $U^\prime$ being adjusted so as to minimize the bilayer thickness deformation energy. As illustrated in Fig.~\ref{fig:12} for a membrane protein with circular protein cross section and constant $\bar U \lambda=0.3$~nm, the value of $U^\prime$ can have a substantial effect on the shape of protein-induced lipid bilayer thickness deformations. In particular, for $U^\prime \approx 0.3$ protein-induced bilayer thickness deformations are seen to decay rapidly in Fig.~\ref{fig:12}.

Plotting $\bar G$ as a function of $U^\prime$ (see Fig.~\ref{fig:13}), we find that $\bar G$ is minimal for $U^\prime \approx 0.28$ with, as suggested by $\bar G_{\mathrm{analy}}$ in Eq.~(\ref{eqGanalytic}), an approximately quadratic dependence of $\bar G$ on $U^\prime$. Allowing for non-circular protein cross sections we find that the optimal $U^\prime$ depends strongly, for large enough $\epsilon$, on the symmetry of clover-leaf protein shapes [see Figs.~\ref{fig:13}(a,b)], but only weakly on the symmetry of polygonal protein shapes [see Fig.~\ref{fig:13}(c)]. Note that, for clover-leaf protein shapes, the optimal $U^\prime$ tend to shift towards $U^\prime \approx 0$ compared to circular protein cross sections. This can be understood by noting that, for clover-leaf protein shapes, the reduction in the size of the membrane footprint brought about by $U \neq 0$ competes with contributions to the bilayer thickness deformation energy due to protein self-interactions. Conversely, polygonal protein shapes only show weak self-interactions, resulting in minor shifts in the optimal $U^\prime$ compared to proteins with circular cross section. Finally, we note that the analytic estimates $\bar G_{\mathrm{analy}}$ in Eq.~(\ref{eqGanalytic}) tend to become less accurate for large $U^\prime$, with up to $\left|\xi_G \right| \approx 60\%$ for the clover-leaf and polygonal shapes considered here [Fig.~\ref{fig:13}(insets)].

\subsection{Variations in protein hydrophobic thickness}
\label{secShapeU}

\begin{figure*}[t!]{\includegraphics{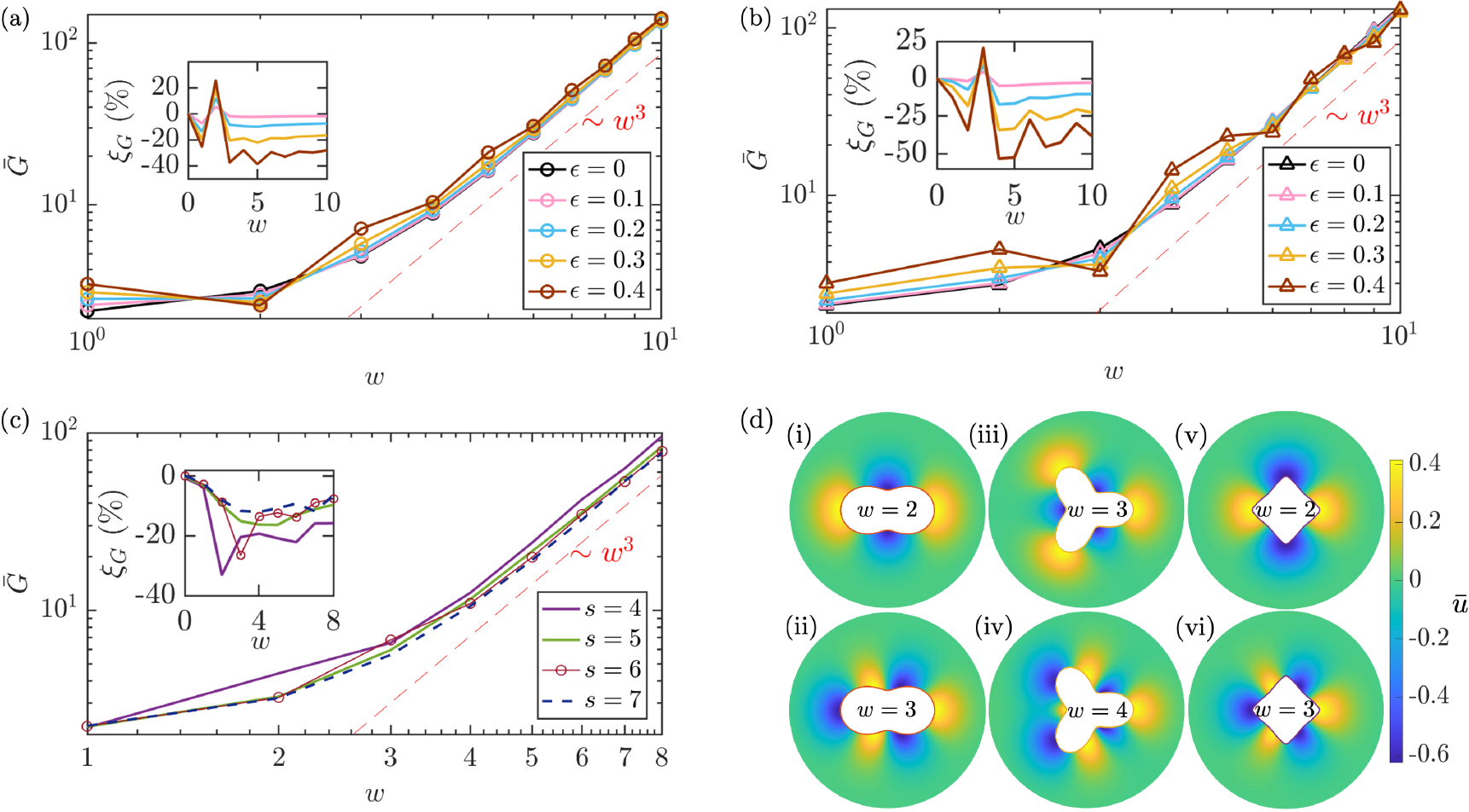}}
	\caption{Bilayer thickness deformation energy $\bar G$ in Eq.~(\ref{eqGAreaDL}) calculated using the BVM (see Sec.~\ref{secBVM}) as a function of the periodicity in protein hydrophobic thickness, $w$ in Eq.~(\ref{eqUBC}), for (a) the clover-leaf protein shapes in Eq.~(\ref{eqCdefclover}) with $s=2$ and the indicated values of $\epsilon$, (b) the clover-leaf protein shapes in Eq.~(\ref{eqCdefclover}) with $s=3$ and the indicated values of $\epsilon$, and (c) the polygonal protein shapes in Eq.~(\ref{eqDefCpoly}) with the indicated values of $s$ and $P=5$. For all panels we set $\bar R \lambda=2.3$~nm, $\bar U_{0} \lambda=-0.1$~nm, $\bar \beta \lambda=0.5$~nm, and $\bar U'=0$. The red dashed lines indicate the asymptotic scaling $\sim w^{3}$. The insets show the signed percent error $\xi_{G}$ in Eq.~(\ref{eqxiG}) for the corresponding analytic approximations $\bar G_{\text{analy}}$ in Eq.~(\ref{eqGanalytic}). In panel (d) we show color maps of the protein-induced bilayer thickness deformations associated with $\epsilon=0.4$ in panel (a) at (i) $w=2$ and (ii) $w=3$, with $\epsilon=0.4$ in panel (b) at (iii) $w=3$ and (iv) $w=4$, and with $s=4$ in panel (c) at (v) $w=2$ and (vi) $w=3$.
	}
	\label{fig:14}
\end{figure*}

Membrane proteins are, in general, expected to show variations in their hydrophobic thickness along the bilayer-protein boundary \cite{Krepkiy2009,Sonntag2011}. For oligomeric membrane proteins, variations in protein hydrophobic thickness are expected to be periodic so as to reflect the protein symmetry. We employ here the sinusoidal variations of $U(\theta)$ in Eq.~(\ref{eqUBC}) as a generic model of variations in protein hydrophobic thickness, in which we denote the periodicity of $U(\theta)$ by $w$. We focus, for now, on zero bilayer-protein contact slopes, $U^\prime = 0$ in Eq.~(\ref{eqUpBC}), but return to the effects of angular variations in $U^\prime$ in Sec.~\ref{secShapeUp}.

Figure~\ref{fig:14} shows that variations in $U(\theta)$ can have a strong impact on the energy cost of protein-induced bilayer thickness deformations, for non-circular as well as circular protein cross sections. The analytic estimate $\bar G_{\mathrm{analy}}$ in Eq.~(\ref{eqGanalytic}) is seen to approximately capture $\bar G$ in Fig.~\ref{fig:14}, but tends to become less accurate as the protein cross section exhibits greater deviations from a circular shape, with up to $\left|\xi_G \right| \approx 50\%$ for the clover-leaf and polygonal protein shapes considered here [Fig.~\ref{fig:14}(insets)]. Note that, for large enough $w$, $\bar G$ in Fig.~\ref{fig:14} scales approximately as $w^3$ for all protein cross sections considered. This can be understood from $\bar G_{\mathrm{analy}}$ in Eq.~(\ref{eqGanalytic}) by noting that $\bar E_{w} \sim w^{3}$ at large $w$, and $\bar \gamma = 0$ if $U^\prime =0$.

While, broadly speaking, variations in protein hydrophobic thickness are seen to increase $\bar G$ in Fig.~\ref{fig:14} for all protein cross sections considered, the interplay of $U(\theta)$ and the shape of the protein cross section can yield comparatively favorable or unfavorable scenarios. For instance, depending on whether adjacent regions of the bilayer-protein boundaries in clover-leaf protein shapes yield bilayer thickness deformations of the same sign [see panels (i) and (iii) in Fig.~\ref{fig:14}(d)] or distinct signs [see panels (ii) and (iv) in Fig.~\ref{fig:14}(d)], protein self-interactions can decrease or increase the energy of protein-induced bilayer thickness deformations. For polygonal protein shapes, we find that scenarios for which the maxima or minima of $U(\theta)$ coincide with the corners of the polygonal shapes [see panel (v) in Fig.~\ref{fig:14}(d)] tend to be unfavorable from an energetic perspective, as compared to scenarios for which the extrema of $U(\theta)$ tend to occur along the polygonal faces [see panel (vi) in Fig.~\ref{fig:14}(d)]. However, compared to the clover-leaf protein shapes considered in Fig.~\ref{fig:14}, the bilayer thickness deformation energy associated with the polygonal protein shapes in Fig.~\ref{fig:14} depends only weakly on the interplay between $U(\theta)$ and the shape of the protein cross section.

\subsection{Variations in bilayer-protein contact slope}
\label{secShapeUp}

\begin{figure*}[t!]{\includegraphics{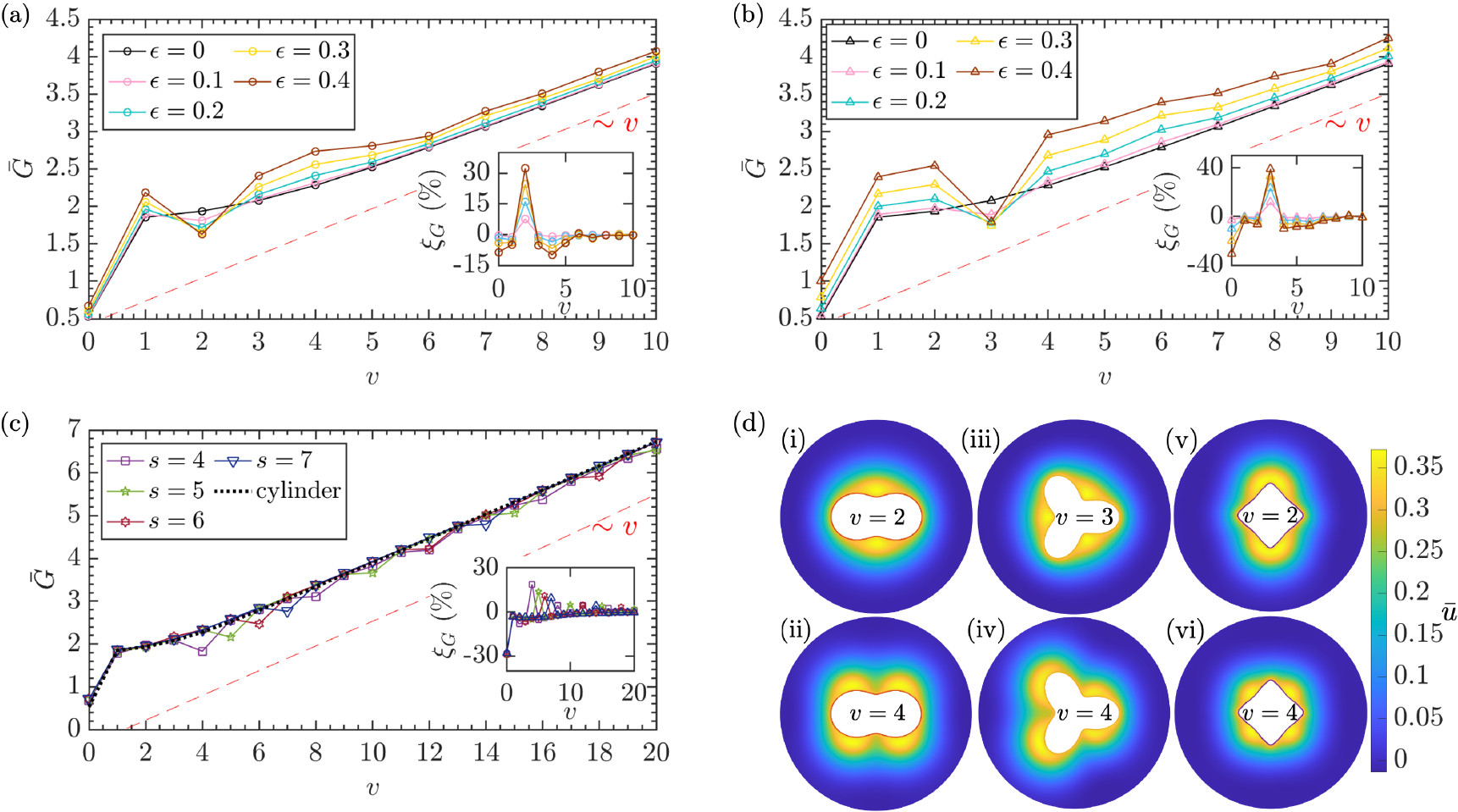}}
	\caption{Bilayer thickness deformation energy $\bar G$ in Eq.~(\ref{eqGAreaDL}) calculated using the BVM (see Sec.~\ref{secBVM}) as a function of the periodicity in the bilayer-protein contact slope, $v$ in Eq.~(\ref{eqUpBC}), for (a) the clover-leaf protein shapes in Eq.~(\ref{eqCdefclover}) with $s=2$ and the indicated values of $\epsilon$, (b) the clover-leaf protein shapes in Eq.~(\ref{eqCdefclover}) with $s=3$ and the indicated values of $\epsilon$, and (c) the polygonal protein shapes in Eq.~(\ref{eqDefCpoly}) with the indicated values of $s$ and $P=5$. For all panels we set $\bar R \lambda=2.3$~nm, $\bar U \lambda =0.3$~nm, $\bar U'_{0}=0$, and $\bar \gamma =0.3$. The red dashed lines indicate the asymptotic scaling $\sim v$. The insets show the signed percent error $\xi_{G}$ in Eq.~(\ref{eqxiG}) for the corresponding analytic approximations $\bar G_{\text{analy}}$ in Eq.~(\ref{eqGanalytic}). In panel (d) we show color maps of the protein-induced bilayer thickness deformations associated with $\epsilon=0.4$ in panel (a) at (i) $v=2$ and (ii) $v=4$, with $\epsilon=0.4$ in panel (b) at (iii) $v=3$ and (iv) $v=4$, and with $s=4$ in panel (c) at (v) $v=2$ and (vi) $v=4$. 
	}
	\label{fig:15}
\end{figure*}

Similarly as the variations in $U(\theta)$ considered in Sec.~\ref{secShapeU}, $U^\prime(\theta)$ in Eq.~(\ref{eqUpBC}) will generally vary along the bilayer-protein interface. Such variations could come about, for instance, through the protein structure or the binding of peptides to some sections of the bilayer-protein interface \cite{Phillips_Ursell_Wiggins_Sens_2009,suchyna04}. Alternatively, if the (normal) gradient of $\bar u$ obeys natural boundary conditions at the bilayer-protein interface, a non-circular protein cross section or variations in $U(\theta)$ may effectively induce variations in $U^\prime(\theta)$. We employ here the simple model of $U^\prime(\theta)$ in Eq.~(\ref{eqUpBC}) to explore the effect of variations in $U^\prime(\theta)$ on the energy cost of protein-induced lipid bilayer thickness deformations. For simplicity we thereby use a constant~$\bar U > 0$.

Figure~\ref{fig:15} illustrates the impact of variations in $U^\prime(\theta)$ on the energy cost of protein-induced lipid bilayer thickness deformations. Similarly as in Fig.~\ref{fig:14}, the analytic estimate $\bar G_{\mathrm{analy}}$ in Eq.~(\ref{eqGanalytic}) is seen to approximately capture $\bar G$ in Fig.~\ref{fig:15}, but tends to become less accurate with increasing deviation of the protein cross section from a circular shape, with up to $\left|\xi_G \right| \approx 40\%$ for the clover-leaf and polygonal shapes considered here [Fig.~\ref{fig:15}(insets)]. Note that, for large enough $v$, $\bar G$ in Fig.~\ref{fig:14} scales approximately linearly with $v$, independent of the protein cross section considered. This can be understood from $\bar G_{\mathrm{analy}}$ in Eq.~(\ref{eqGanalytic}) by noting that $\bar F_{v} \sim v$ at large $v$. In analogy to Fig.~\ref{fig:14} we find in Fig.~\ref{fig:15} that, broadly speaking, variations in $U^\prime(\theta)$ increase $\bar G$ for all protein cross sections considered. However, the interplay of $U^\prime(\theta)$ and the shape of the protein cross section can yield, similarly as in Fig.~\ref{fig:14}, comparatively favorable or unfavorable scenarios. In particular, for the clover-leaf protein shapes in Figs.~\ref{fig:15}(a) and~\ref{fig:15}(b), it tends to be energetically favorable for the minima of $\bar U'(\theta)$ to coincide with the minima of $\bar C(\theta)$, so as to make protein self-interactions more favorable, and the maxima of $\bar U'(\theta)$ to coincide with the maxima of $\bar C(\theta)$, so as to reduce the protein's membrane footprint. This configuration is achieved, for instance, when $v=s$ [see panels (i) and (iii) in Fig.~\ref{fig:15}(d)]. Conversely, it tends to be energetically unfavorable for the minima of $\bar U'(\theta)$ to coincide with the maxima of $\bar C(\theta)$, and vice versa, or for $\bar U'(\theta)$ and $\bar C(\theta)$ to be out of phase [see panels (ii) and (iv) in Fig.~\ref{fig:15}(d)]. For the polygonal protein shapes in Fig.~\ref{fig:15}(c), particularly favorable configurations tend to be achieved when the minima of $\bar U'(\theta)$ fall on the polygonal faces, rather than on the corners of the polygonal shapes [see panels (v) and (vi) in~Fig.~\ref{fig:15}(d)].

\section{Transitions in protein organization and shape}
\label{secStability}

Section~\ref{secDependShape} demonstrates that protein-induced lipid bilayer thickness deformations show a strong dependence on protein shape, and that changes in protein shape can bring about changes in the bilayer thickness deformation energy $>10$~$k_B T$ in magnitude. In the present section we suggest possible implications of these results for the biophysical properties of membrane proteins. In particular, Sec.~\ref{secStabilityOligomers} explores the energetic contribution of lipid bilayer thickness deformations to the self-assembly of protein monomers into protein oligomers, and how changes in bilayer-protein interactions could destabilize protein oligomers. In Sec.~\ref{secStabilityShape} we investigate the effect of lipid bilayer thickness deformations on transitions in protein conformational state that involve substantial changes in protein shape.

\subsection{Self-assembly of protein oligomers}
\label{secStabilityOligomers}

\begin{figure}{\includegraphics{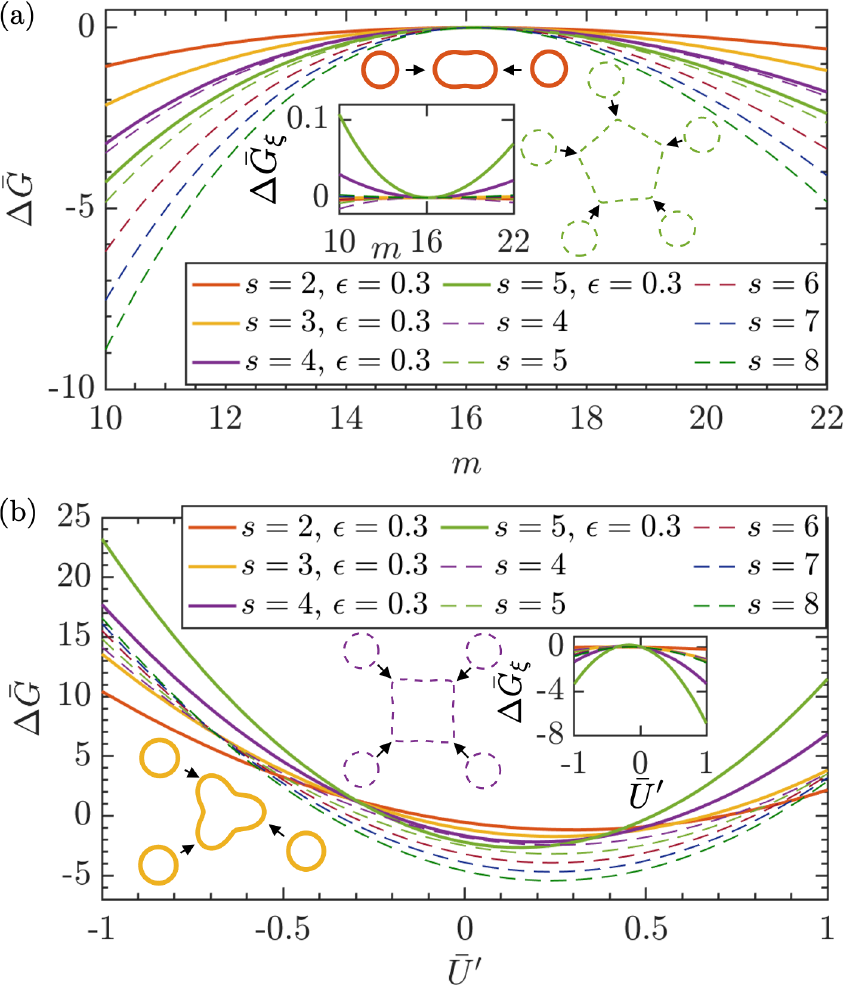}}
	\caption{Difference between the lipid bilayer thickness deformation energies associated with protein oligomers of symmetry $s$ and their corresponding $s$ monomers, $\Delta \bar G$, calculated using the BVM (see Sec.~\ref{secBVM}) as a function of (a) the lipid chain length $m$ in Eq.~(\ref{eqDefm}) and (b) the (constant) bilayer-oligomer contact slope $U^\prime$ in Eq.~(\ref{eqDefUp}) for a variety of clover-leaf (solid curves) and polygonal (dashed curves) shapes of the protein oligomers. We took the protein monomers to have circular cross sections with $U^\prime = 0$ and used the indicated values of $s$, with $\epsilon = 0.3$ for the clover-leaf oligomer shapes in Eq.~(\ref{eqCdefclover}) and $P=5$ for the polygonal oligomer shapes in Eq.~(\ref{eqDefCpoly}). We set $\bar R \lambda = 1$~nm for the monomer radii, and used identical cross-sectional areas of the oligomers and their corresponding monomers. We set $U^\prime = 0$ in panel (a), $2\bar a\lambda = 3.2$~nm in panel (b), and used $\bar W \lambda=3.8$~nm for the protein monomers and oligomers in all panels. The schematics in the insets illustrate transitions between monomers and oligomers for selected oligomeric shapes. The plots in the insets show the difference in the oligomerization energies obtained from the analytic approximation $\bar G_{\mathrm{analy}}$ in Eq.~(\ref{eqGanalytic}) and the BVM, $\Delta\bar G_{\xi}=\Delta \bar G_{\mathrm{analy}}-\Delta \bar G$, for each curve in the main~panels.}
	\label{fig:16}
\end{figure}

Complex molecular architectures of membrane proteins often arise from self-assembly of small protein subunits (monomers) into protein oligomers \cite{vinothkumar2010structures,forrest2015structural}. The thermodynamic competition between different oligomeric states of membrane proteins depends crucially on how the energy per protein subunit changes with protein oligomeric state \cite{kahraman2016thermodynamic}. If the hydrophobic thickness of the protein oligomers or monomers differs from the unperturbed hydrophobic thickness of the surrounding lipid bilayer, one set of contributions to the oligomerization energy arises from protein-induced bilayer thickness deformations. We illustrate here how these contributions to the oligomerization energy can be calculated through the BVM for protein-induced bilayer thickness deformations. For simplicity, we thereby consider a protein oligomer of symmetry $s$ with a clover-leaf or polygonal cross section, and take the $s$ (identical) competing protein monomers to have circular cross sections with the same total area as the protein oligomer and no interactions between the monomers. Furthermore, we assume that the protein oligomers and monomers show constant values of $U$ and $U^\prime$ along the bilayer-protein interface, with identical $U$ for the protein oligomers and monomers and $U^\prime = 0$ for the protein monomers. These assumptions could easily be lifted to describe more complex scenarios.

Figure~\ref{fig:16} shows the difference between the (dimensionless) bilayer thickness deformation energies associated with protein oligomers and their corresponding monomers, $\Delta \bar G$, as a function of the lipid chain length $m$ [see Fig.~\ref{fig:16}(a)] and the bilayer-oligomer contact slope $U^\prime$ [see Fig.~\ref{fig:16}(b)] for a variety of shapes of the oligomer cross section. The insets in Fig.~\ref{fig:16} show the differences in the oligomerization energies obtained from the analytic approximation $\bar G_{\mathrm{analy}}$ in Eq.~(\ref{eqGanalytic}) and the BVM, $\Delta\bar G_{\xi}=\Delta \bar G_{\mathrm{analy}} - \Delta \bar G$. Equation~(\ref{eqGanalytic}) is seen to provide, for modest magnitudes of $U$ and $U^\prime$, good estimates of the oligomerization energy. We generally have $\Delta G < 0$ in Fig.~\ref{fig:16}(a), indicating that protein-induced lipid bilayer thickness deformations support oligomerization. This can be understood from $\bar G_{\mathrm{analy}}$ by noting that the protein oligomers in Fig.~\ref{fig:16}(a) have a smaller circumference than their corresponding monomers. Interestingly, Fig.~\ref{fig:16}(b) shows that $\Delta G$ can become positive for large enough magnitudes of $U^\prime$ for the protein oligomer, thus destabilizing the protein oligomer. Such a change in $U^\prime$ could be achieved, for instance, through a transition in the conformational state of the oligomer or the binding of peptides to the oligomer \cite{Phillips_Ursell_Wiggins_Sens_2009,suchyna04}. Figure~\ref{fig:16} thus suggests that protein-induced bilayer thickness deformations could assist both in the assembly and disassembly of protein oligomers, and contribute $>10$~$k_B T$ to the energy budget of oligomer assembly or disassembly.

\subsection{Transitions in protein conformational state}
\label{secStabilityShape}

\begin{figure*}[t!]{\includegraphics{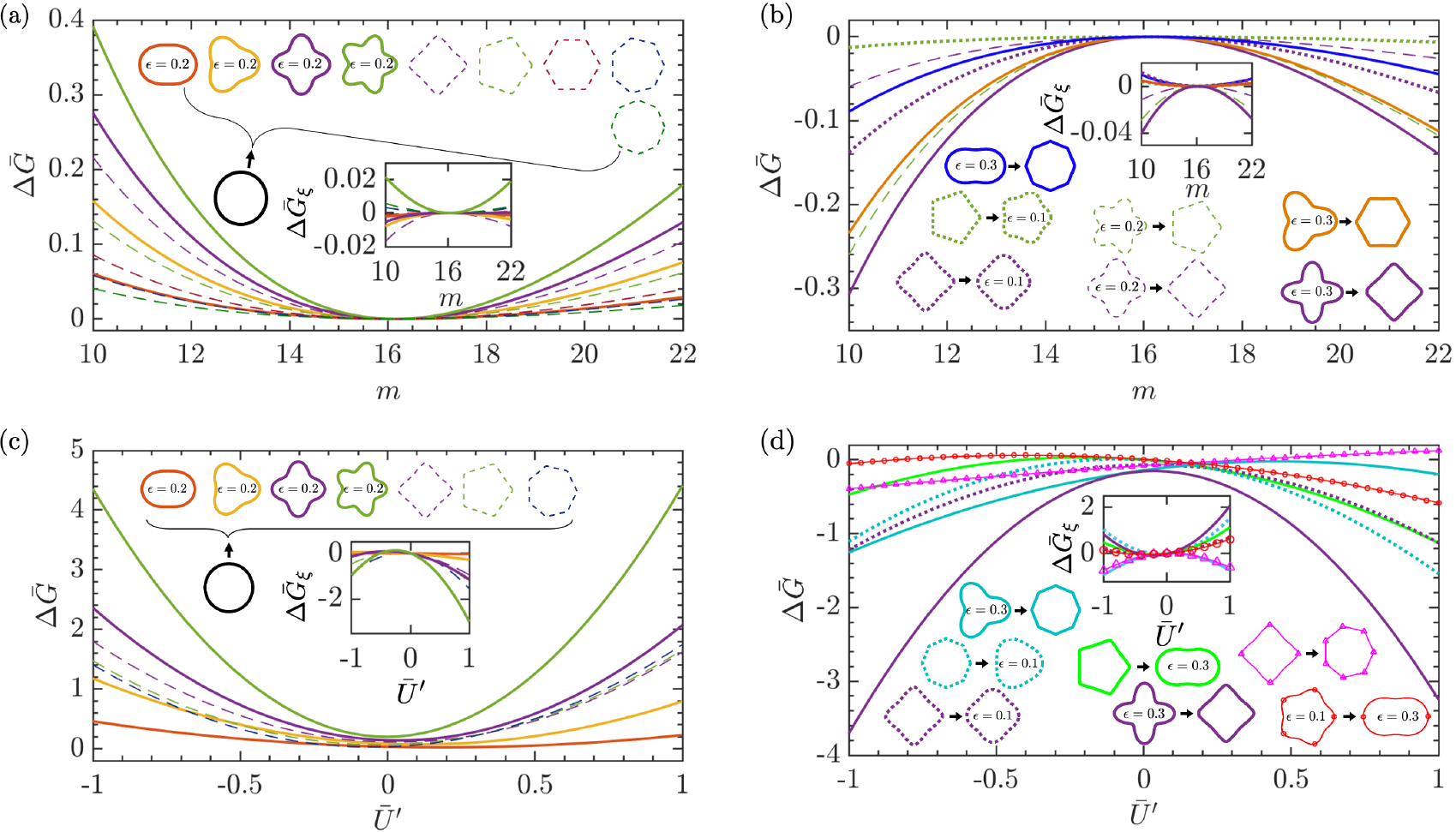}}
	\caption{Difference between the lipid bilayer thickness deformation energies associated with the final and initial protein shapes indicated in the insets, $\Delta \bar G$, calculated using the BVM (see Sec.~\ref{secBVM}) as a function of (a,b) the lipid chain length $m$ in Eq.~(\ref{eqDefm}) and (c,d) the (constant) bilayer-oligomer contact slope $U^\prime$ in Eq.~(\ref{eqDefUp}). The values of $\epsilon$ associated with each clover-leaf shape in Eq.~(\ref{eqCdefclover}) are indicated in the insets, while for the polygonal protein shapes we set $P=5$. We set $U^\prime = 0$ in panels (a,b) and $2\bar a\lambda=3.2$~nm in panels (c,d), and used $\bar W \lambda=3.8$~nm for all panels. The cross sections of all protein shapes considered here have area $\pi \bar R^{2}$ with $\bar R \lambda= 2.3$~nm. The plots in the insets show the differences in the protein transition energies obtained from the analytic approximation $\bar G_{\mathrm{analy}}$ in Eq.~(\ref{eqGanalytic}) and the BVM, $\Delta\bar G_{\xi}=\Delta \bar G_{\mathrm{analy}} - \Delta \bar G$, for each curve in the main panels.}
	\label{fig:17}
\end{figure*}

To perform their biological functions, membrane proteins often have to transition between different conformational states. Such transitions in protein conformational state can be accompanied by changes in the cross-sectional shape of membrane proteins producing, in turn, changes in protein-induced lipid bilayer deformations. Membrane proteins can thus be regulated by lipid bilayer properties, such as the bilayer hydrophobic thickness \cite{Perozo_Kloda_Cortes_Martinac_2002,rusinova2021mechanisms,Phillips_Ursell_Wiggins_Sens_2009,Andersen_Koeppe_2007}. We illustrate here how the BVM can be used to calculate the contribution of lipid bilayer thickness deformations to the energy difference between two protein states with distinct cross-sectional shapes. For simplicity, we thereby take the two states of the membrane protein to show identical $U$ and $U^\prime$ that are constant along the bilayer-protein interface, and to have cross-sectional shapes with the same area. These assumptions could easily be lifted to provide detailed models of specific conformational transitions in membrane proteins, which may also involve more than just two protein states.

Figure~\ref{fig:17} shows the difference between the lipid bilayer thickness deformation energies associated with the final and initial protein shapes indicated in the insets, $\Delta \bar G$, as a function of the lipid chain length $m$ [see Figs.~\ref{fig:17}(a) and~\ref{fig:17}(b)] and the bilayer-protein contact slope $U^\prime$ [see Figs.~\ref{fig:17}(c) and~\ref{fig:17}(d)]. The insets in Fig.~\ref{fig:17} show the corresponding differences in the protein transition energies obtained from the analytic approximation $\bar G_{\mathrm{analy}}$ in Eq.~(\ref{eqGanalytic}) and the BVM, $\Delta\bar G_{\xi}=\Delta \bar G_{\mathrm{analy}} - \Delta \bar G$. Similarly as in Fig.~\ref{fig:16}, Eq.~(\ref{eqGanalytic}) is seen to provide, for modest magnitudes of $U$ and $U^\prime$, good estimates of $\Delta \bar G$ in Fig.~\ref{fig:17}. In Figs.~\ref{fig:17}(a) and~\ref{fig:17}(c) we consider idealized scenarios in which the initial protein shape shows a circular cross section, while the final state corresponds to a clover-leaf or polygonal shape. We find that bilayer thickness deformations generally inhibit such transitions in protein shape, $\Delta \bar G \geq 0$, which is easily understood from $\bar G_{\mathrm{analy}}$ in Eq.~(\ref{eqGanalytic}) by noting that these transitions in protein shape are accompanied by an increase in protein circumference. In Fig.~\ref{fig:17}(b) we study $\Delta \bar G$ for transitions between proteins with non-circular cross sections. We thereby arranged the initial and final protein states such that $\Delta \bar G \leq 0$. Similarly as in Figs.~\ref{fig:17}(a) and~\ref{fig:17}(c), the results in Fig.~\ref{fig:17}(b) can be understood by noting that the transitions in protein shape in Fig.~\ref{fig:17}(b) are accompanied by a decrease in protein circumference. Note, in particular, that the energetically favorable protein shapes in Fig.~\ref{fig:17}(b) tend to correspond to polygonal protein shapes or clover-leaf shapes with small $\epsilon$.

Finally, we consider in Fig.~\ref{fig:17}(d) scenarios where the sign of $\Delta \bar G$ does not necessarily follow from the relative protein circumferences of the initial and final protein shapes, and may not be captured by $G_{\mathrm{analy}}$ in Eq.~(\ref{eqGanalytic}) for all the values of $U$ and $U^\prime$ considered here. In particular, we chose the initial and final protein shapes associated with the dotted teal, green, and red curves in Fig.~\ref{fig:17}(d) so that their circumferences lie within $1\%$ of each other, and the remaining protein shapes so that the final protein shape has a circumference that is substantially smaller than that of the initial state, by at least $6\%$. The former sets of protein shapes yield a change of sign in $\Delta \bar G$ with $U^\prime$. Three of the latter sets of protein shapes, corresponding to the teal, purple, and dotted purple curves in Fig.~\ref{fig:17}(d), always yield $\Delta \bar G \leq 0$, which can again be understood from $\bar G_{\mathrm{analy}}$ in Eq.~(\ref{eqGanalytic}), while the fourth, corresponding to the pink curve in Fig.~\ref{fig:17}(d), can yield a change of sign in $\Delta \bar G$ with $U^\prime$. In analogy to Fig.~\ref{fig:16}(b), this suggests that, for suitable protein shapes, modification of $U^\prime$ in a given (stable) protein conformational state through, for instance, peptide binding \cite{Phillips_Ursell_Wiggins_Sens_2009,suchyna04} could trigger a change in the protein conformational state mediated by bilayer thickness deformations. We note, however, that for the protein shapes in Fig.~\ref{fig:17}(d) $\Delta G$ exceeds zero by not more than a few $k_B T$.

\section{Summary and conclusions}
\label{secConclusion}

Employing protein-induced lipid bilayer thickness deformations as a test case \cite{Andersen_Koeppe_2007,Phillips_Ursell_Wiggins_Sens_2009,Huang_1986,Dan_Pincus_Safran_1993,Dan_Berman_Pincus_Safran_1994,Wiggins_Phillips_2005,Ursell_Kondev_Reeves_Wiggins_Phillips_2008,mondal11,Kahraman_Koch_Klug_Haselwandter_PRE_2016,Argudo_Bethel_Marcoline_Wolgemuth_Grabe_2017,canham70,Helfrich_1973,evans74,zimmerberg06,weikl18,young22,fournier99,Rawicz_Olbrich_McIntosh_Needham_Evans_2000}, we have described here a BVM that permits the construction of non-perturbative analytic solutions of protein-induced lipid bilayer deformations for non-circular membrane protein cross sections. In addition to the membrane protein cross section, our BVM allows for a breaking of rotational symmetry about the protein center through angular variations in the protein hydrophobic thickness and the bilayer-protein contact slope along the bilayer-protein interface. Our BVM reproduces available analytic solutions for membrane proteins with circular cross section \cite{Huang_1986,Nielsen_Goulian_Andersen_1998,Wiggins_Phillips_2005,Haselwandter_Phillips_PLOS_2013,Haselwandter_Phillips_EPL_2013,Kahraman_Koch_Klug_Haselwandter_PRE_2016} and yields, for membrane proteins with non-circular cross section, excellent agreement with numerical, finite element solutions. Based on these BVM solutions, we formulated a simple analytic approximation of the lipid bilayer thickness deformation energy associated with general protein shapes [see Eq.~(\ref{eqGanalytic}) with Eq.~(\ref{eqDEFH})]. We find that, for modest deviations from rotational symmetry, this analytic approximation is in good agreement with BVM solutions. These results suggest that, to a first approximation, the effect of membrane protein shape on the energy of bilayer thickness deformations can be understood in terms of the circumference associated with non-circular protein cross sections.

Through our BVM and analytic approximation of the lipid bilayer thickness deformation energy, we surveyed the dependence of protein-induced lipid bilayer thickness deformations on protein shape. We find that protein shape tends to have a large effect on the energy of protein-induced lipid bilayer thickness deformations, typically shifting the bilayer deformation energy by more than 10~$k_B T$. A limitation of the BVM described here arises for protein shapes that show extreme deviations from circular symmetry, in which case BVM solutions tend to involve a large number of terms and, hence, become increasingly intractable. In such cases it may be advisable to modify the APD method for the distribution of boundary points employed here, so as to reduce the number of terms required in the lipid bilayer thickness deformation field in Eq.~(\ref{equEigenSuperN}) with Eq.~(\ref{eqEigenFuncsN}). While we have focused here on bilayer thickness deformations, an approach analogous to the BVM employed here could be used to construct non-perturbative analytic solutions for other modes of membrane deformation, such as bilayer midplane deformations \cite{canham70,Helfrich_1973,evans74,zimmerberg06,weikl18,young22}. Furthermore, it would be interesting to construct BVM solutions for membrane proteins embedded in bilayers with heterogeneous lipid composition \cite{Leibler1987,SensSafran2000,Schaffer2004,Shrestha_Kahraman_Haselwandter_2020,shrestha22}.

We find that protein self-interactions provide an important motif for the energy of protein-induced lipid bilayer thickness deformations. Such self-interactions arise for invaginations in the protein cross section, from overlaps in the bilayer deformations induced at different sections of the bilayer-protein interface. The basic phenomenology of membrane protein self-interactions can be understood by drawing analogies to bilayer-thickness-mediated interactions between proteins \cite{Dan_Pincus_Safran_1993,Dan_Berman_Pincus_Safran_1994,Ursell2007,harroun99,goforth03,botelho06,Phillips_Ursell_Wiggins_Sens_2009,grage11,Haselwandter_Phillips_EPL_2013,Kahraman_Klug_Haselwandter_2014,Haselwandter_Wingreen_2014,milovanovic15,Kahraman_Koch_Klug_Haselwandter_PRE_2016,Kahraman_Koch_Klug_Haselwandter_SR_2016,pollard18}. In particular, membrane protein self-interactions can effectively lower the energy cost of protein-induced lipid bilayer thickness deformations for proteins with constant bilayer-protein hydrophobic mismatch and zero bilayer-protein contact slope. For non-zero bilayer-protein contact slopes, or for variations in the bilayer-protein hydrophobic mismatch or in the bilayer-protein contact slope along the bilayer-protein interface, protein self-interactions can yield dramatic shifts in the bilayer thickness deformation energy. Thus, the interplay between the cross-sectional shape of membrane proteins, protein hydrophobic thickness, and bilayer-protein contact slope yields a rich energy landscape of protein-induced lipid bilayer thickness deformations. Interestingly, the hydrophobic thickness or bilayer-protein contact slope of membrane proteins may be modified in cells through, for instance, protein mutations, changes in lipid composition, or the binding of peptides at the bilayer-protein interface, while protein oligomerization and transitions in protein conformational state tend to modify the cross-sectional shape of membrane proteins. The results described here therefore suggest general physical mechanisms for how protein shape couples to the function, regulation, and organization of membrane proteins.

\acknowledgements

This work was supported by NSF Grants No.~MCB-2202087, No.~DMR-2051681, and No.~DMR-1554716, a USC Graduate School DIA Fellowship, and the USC Center for Advanced Research Computing. We thank O.~Kahraman for helpful discussions on the finite element method, M.~Olguin for useful discussions on code optimization, and F.~Pinaud for helpful discussions on the self-assembly of membrane protein complexes.

\appendix

\section{Computational implementation of the boundary value method}
\label{AppA}

The general solution for the bilayer thickness deformation field in Eq.~(\ref{equEigenSuperN}) with Eq.~(\ref{eqEigenFuncsN}) involves modified Bessel functions of the second kind, $K_{n}$, of any order $n$ \cite{Abramowitz_Stegun_1964}. $K_{n}(x)$ can vary rapidly with $x$, leading to numerical overflow and large round-off (floating point) errors \cite{Atkinson_1988}. These numerical issues are compounded by error propagation in the arithmetic operations necessary for solving the linear system of equations relevant for the BVM \cite{Atkinson_1988,Trefethen_1997}. In particular, if the values of the matrix $\mathbf{A}$ in Eq.~(\ref{eqmeq}) vary over many orders of magnitude, which is typically the case for the scenarios considered here, the resulting propagation of floating point errors can be catastrophic. These problems are ameliorated through the APD method, which effectively reduces the number of terms required in the series in Eq.~(\ref{equEigenSuperN}) with Eq.~(\ref{eqEigenFuncsN}), as well as LU decomposition with partial pivoting of $\mathbf{A}$ \cite{Atkinson_1988,Trefethen_1997}, which reduces the pairing of matrix elements that differ over many orders of magnitude.

We solved the linear system of equations in Eq.~(\ref{eqmeq}) in C++ using F.~Johansson's arbitrary precision library for C/C++, \textit{Arb} \cite{Johansson_arb_2017}, which includes built-in functions for LU decomposition with partial pivoting. Importantly, \textit{Arb} also includes Bessel functions with support for complex arguments. The linear system of equations in Eq.~(\ref{eqmeq}) encompasses $Q=4N+2$ independent equations. As $N$ is increased in Eq.~(\ref{equEigenSuperN}) with Eq.~(\ref{eqEigenFuncsN}), solving Eq.~(\ref{eqmeq}) therefore becomes increasingly intensive from a computational perspective. To improve the computational efficiency of our calculations, we use \textit{OpenMP} multi-threading \cite{dagum_omp_1998} to spread computations across multiple CPU cores.

\begin{figure}[t!]
	{\includegraphics[width=\columnwidth]{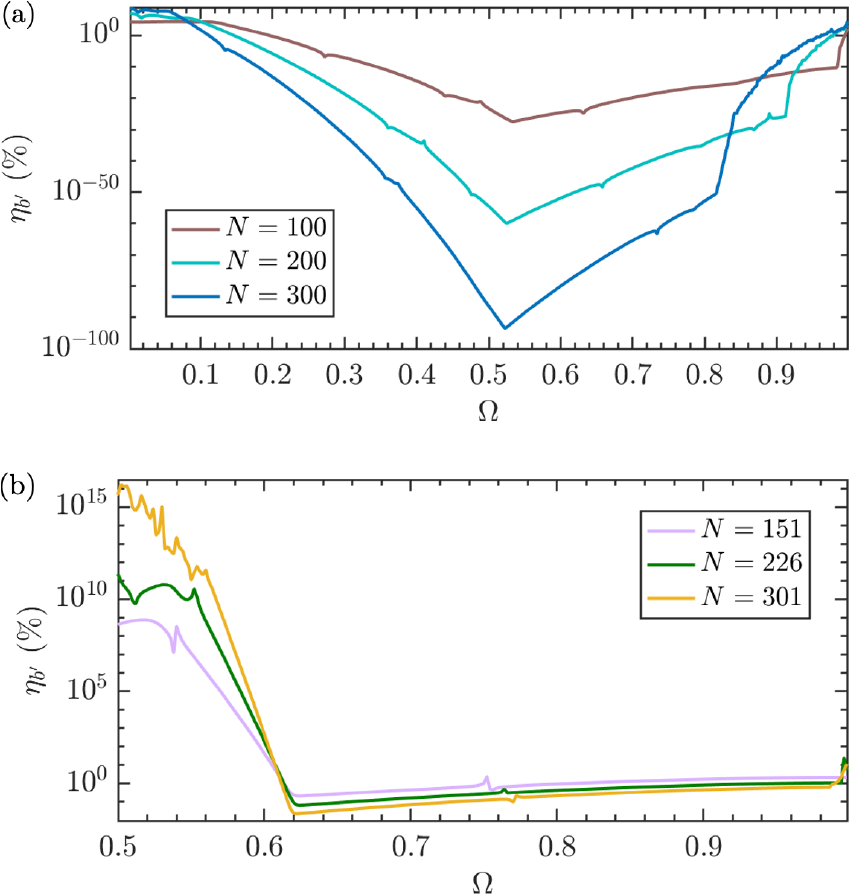}}
	\caption{Boundary error $\eta_{b'}$ in Eq.~(\ref{eqetabdef}) in BVM calculations (see Sec.~\ref{secBVM}) for clover-leaf protein shapes with (a) $s=1$ and $\epsilon=0.54$ in Eq.~(\ref{eqCdefclover}) and (b) $s=3$ and $\epsilon=0.38$ in Eq.~(\ref{eqCdefclover}) as a function of the gap factor $\Omega$ in Eq.~(\ref{eqelldef}). We set $R\approx2.3$~nm, $U=0.3$ nm, and $U'=0$. For ease of comparison we used, for each curve, the indicated, fixed values of $N$ in Eq.~(\ref{equEigenSuperN}) with Eq.~(\ref{eqEigenFuncsN}).} \label{fig:18}
\end{figure}

As discussed in Sec.~\ref{secBVM}, the APD method involves the gap factor $\Omega$ in Eq.~(\ref{eqelldef}), which we optimized so that the boundary error $\eta_{b'} \leq 0.1 \%$ in Eq.~(\ref{eqetabdef}) and we obtained changes in $G$ and $\eta_{b'}$ of no more than $10^{-5}\%$ as the numerical precision was increased. A suitable choice for $\Omega$ thus allows construction of accurate solutions through Eq.~(\ref{equEigenSuperN}) with Eq.~(\ref{eqEigenFuncsN}) at lower orders $N$, thus improving the numerical performance of the BVM. For example, Fig.~\ref{fig:18} shows $\eta_{b'}$ as a function of $\Omega$ for clover-leaf protein shapes with symmetries $s=1$ [see Fig.~\ref{fig:18}(a)] and $s=3$ [see Fig.~\ref{fig:18}(b)], with constant $U$ and $U'$ along the bilayer-protein interface. The solutions in Fig.~\ref{fig:18} were computed at the indicated orders $N$ in Eq.~(\ref{equEigenSuperN}) with Eq.~(\ref{eqEigenFuncsN}). Figure~\ref{fig:18} illustrates how the optimal gap factor $\Omega$ converges with increasing $N$. For clover-leaf protein shapes we generally find that the optimal $\Omega$ increases with increasing $\epsilon$. For the scenarios considered here we also find that, for a given $N$ in Eq.~(\ref{equEigenSuperN}) with Eq.~(\ref{eqEigenFuncsN}) and shape of the protein cross section, the optimal $\Omega$ changes only weakly if one allows for variations in $U$ or $U'$ along the bilayer-protein interface.

\section{Numerical precision}
\label{AppB}

\begin{figure}[b!]
	{\includegraphics[width=\columnwidth]{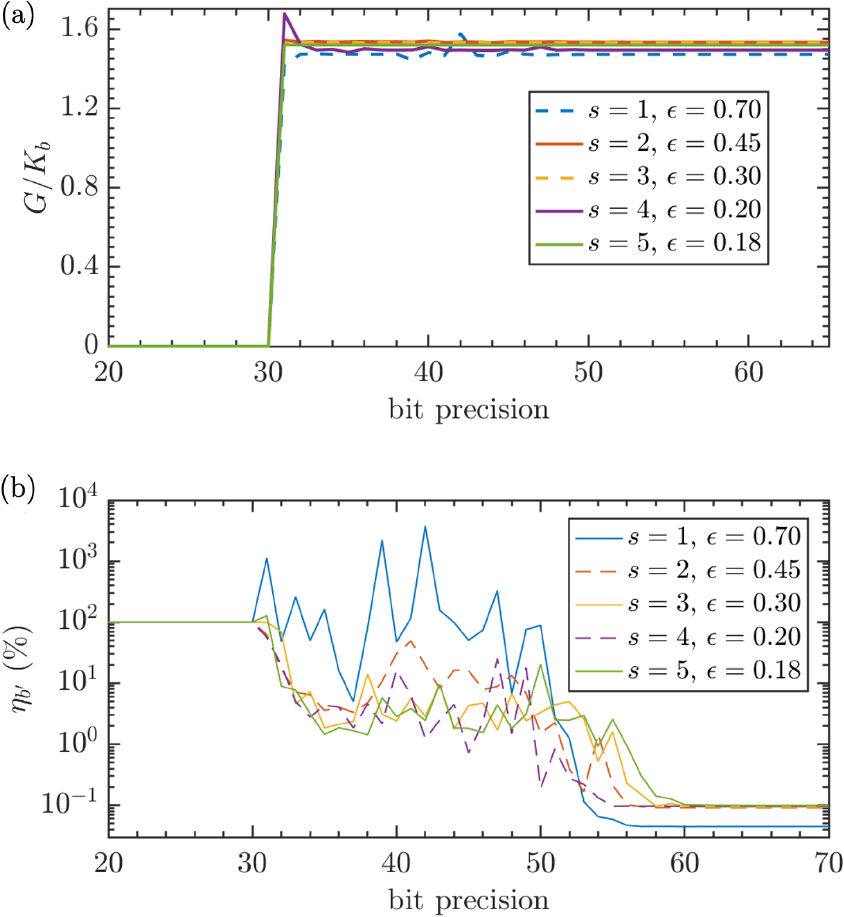}}
	\caption{(a) Lipid bilayer thickness deformation energy $G$ in Eq.~(\ref{eqGBVM}) calculated using the BVM (see Sec.~\ref{secBVM}) and (b) corresponding boundary error $\eta_{b'}$ in Eq.~(\ref{eqetabdef}) for clover-leaf protein shapes as a function of the bit precision employed in the numerical computations. We used the indicated values of $s$ and $\epsilon$ in Eq.~(\ref{eqCdefclover}), and $R\approx2.3$~nm, $U=0.3$~nm, and $U'=0$. For ease of comparison we used for $s=1$ the fixed values $N=20$ in Eq.~(\ref{equEigenSuperN}) with Eq.~(\ref{eqEigenFuncsN}) and $\Omega=0.726$ in Eq.~(\ref{eqelldef}), $N=48$ and $\Omega=0.62$ for $s=2$, $N=72$ and $\Omega=0.52$ for $s=3$, $N=90$ and $\Omega=0.552$ for $s=4$, and $N=125$ and $\Omega=0.45$ for $s=5$.}\label{fig:19}
\end{figure}

For the numerical calculations of the lipid bilayer thickness deformation energy $G$ in Eq.~(\ref{eqGBVM}) presented in this article, we generally used numbers with precision (substantially) greater than double precision (64 bits) \cite{Johansson_arb_2017}, so as to meet the numerical precision criteria described in Sec.~\ref{secBVM} with the boundary error $\eta_{b'} \leq 0.1 \%$ in Eq.~(\ref{eqetabdef}) and changes in $G$ and $\eta_{b'}$ of no more than $10^{-5}\%$ as the numerical precision is increased (see also Appendix~\ref{AppA}). However, many programming languages do not have built-in support for numerical precision greater than double precision. To illustrate the extent to which double precision calculations could be used to approximate the BVM results described here, we plot in Fig.~\ref{fig:19} the bilayer thickness deformation energy $G$ [see Fig.~\ref{fig:19}(a)] and the corresponding boundary error $\eta_{b'}$ [see Fig.~\ref{fig:19}(b)] versus bit precision for several clover-leaf protein symmetries $s$ and the indicated values of $\epsilon$. As described in Appendix~\ref{AppA}, the results in Fig.~\ref{fig:19} were obtained with F.~Johansson's arbitrary precision library for C/C++, \textit{Arb} \cite{Johansson_arb_2017}. We have $\eta_{b'} \leq 0.1 \%$ in Fig.~\ref{fig:19} as the floating point precision is increased beyond double precision, with changes in $G$ and $\eta_{b'}$ of no more than $10^{-5}\%$. For the clover-leaf protein shapes considered in this article, we generally find that numerical precision greater than double precision is required for large $s$ or $\epsilon$. For the polygonal protein shapes considered in this article, we find that a numerical precision greater than double precision is generally required to meet the numerical precision criteria described in Sec.~\ref{secBVM}.

\bibliography{Refs}

\end{document}